\documentclass[preprint, nofootinbib, aps, floats, floatfix, amsmath, amssymb, longbibliography, superscriptaddress, preprintnumbers]{revtex4-1} %
\pdfoutput=1

\usepackage[final]{graphicx}
\usepackage{amsmath}
\usepackage{ulem}
\usepackage{bbm}
\usepackage{bm}
\usepackage{amsfonts}
\usepackage{amssymb}
\usepackage{latexsym}
\usepackage{graphicx}
\usepackage[english]{babel}
\usepackage{multirow}
\usepackage{float}
\usepackage{url}
\usepackage{slashed}
\usepackage{xcolor} 
\usepackage[utf8]{inputenc}
\usepackage{stmaryrd} 
\usepackage{enumitem}
\usepackage{hyperref}
\usepackage{cleveref}
\usepackage{siunitx}
\usepackage{verbatim}
\usepackage{epstopdf}
\usepackage{amsthm}
\usepackage{subfigure}

\begin{document}
\title{The ALP explanation to muon $(g-2)$ \\ and its test at future Tera-$Z$ and Higgs factories }

\author{Jia Liu}
\email{jialiu@pku.edu.cn}
\affiliation{School of Physics and State Key Laboratory of Nuclear Physics and Technology, Peking University, Beijing 100871, China}
\affiliation{Center for High Energy Physics, Peking University, Beijing 100871, China}

\author{Xiaolin Ma}
\email{themapku@stu.pku.edu.cn}
\affiliation{School of Physics and State Key Laboratory of Nuclear Physics and Technology, Peking University, Beijing 100871, China}

\author{Lian-Tao Wang}
\email{liantaow@uchicago.edu}
\affiliation{Enrico Fermi Institute and Department of Physics, The University of Chicago, 
		Chicago, IL 60637, USA}
\affiliation{Kavli Institute for Cosmological Physics, 
		The University of Chicago, Chicago, IL 60637, USA}

\author{Xiao-Ping Wang}
\email{hcwangxiaoping@buaa.edu.cn}
\affiliation{School of Physics, Beihang University, Beijing 100083, China}
\affiliation{Beijing Key Laboratory of Advanced Nuclear Materials and Physics, Beihang University, Beijing 100191, China}	
	
\date{\today}
	
\begin{abstract}
Models with an Axion Like Particle (ALP) can provide an explanation for the discrepancy between experimental measurement of the muon anomalous magnetic moment $(g-2)_\mu$ and the Standard Model prediction. This explanation relies on the couplings of the ALP to the muon and the photon. We also include more general couplings to the electroweak gauge bosons and incorporate them in the calculations up to the 2-loop order. 
We investigate the existing experimental constraints and find that they do not rule out the ALP model under consideration as a possible explanation for the  $(g-2)_\mu$ anomaly. At the same time, we find the future Tera-$Z$ and Higgs factories, such as the CEPC and FCC-ee, can completely cover the relevant parameter space through searches with final states $(\gamma \gamma) \gamma$, $(\mu^+ \mu^-) \gamma$ and $(\mu^+ \mu^-) \mu^+ \mu^-$.
\end{abstract}
\maketitle
\tableofcontents

\section{Introduction}
The Strong CP problem in the Standard Model (SM) \cite{Crewther:1979pi,Pignol:2021uuy,Golub:1994cg,Abel:2020pzs} can be solved by the Peccei-Quinn mechanism, leading to the prediction of the existence a pseudo Nambu-Goldstone boson,  the axion~\cite{Peccei:1977hh, Peccei:1977ur, Weinberg:1977ma, Wilczek:1977pj,Vafa:1984xg}. 
The shift symmetry of the axion implies that it only has derivative couplings except for non-perturbative effects through strong interactions. In addition, other axion-like particles (ALPs) are ubiquitous in many new physics scenarios. Similar to the QCD axion, an ALP is a pseudo-Goldstone boson with an approximate shift symmetry. The corresponding decay constant $f_a$ and mass $m_a$ can be free parameters. It shares similar interactions with the QCD axion at low energy and predicts rich phenomenology to be explored in various experiments \cite{Raffelt:1990yz, Duffy:2009ig, Kawasaki:2013ae, Marsh:2015xka, Graham:2015ouw, Bauer:2017ris, DiLuzio:2020wdo}.

The anomalous magnetic dipole moment of the muon, $(g-2)_\mu$ is a powerful tool to test the SM and probe the physics beyond the Standard Model~\cite{ACME:2018yjb, Muong-2:2008ebm, Hanneke:2008tm, Hanneke:2010au, Aoyama:2020ynm}. The combined results of the measurements at the Brookhaven National Laboratory~\cite{Muong-2:2006rrc} and the Fermi National Accelerator Laboratory~\cite{Muong-2:2021ojo} suggest a 4.2 $\sigma$ discrepancy between SM prediction~\cite{Davier:2010nc, Aoyama:2012wk, Aoyama:2019ryr, Czarnecki:2002nt, Gnendiger:2013pva, Davier:2017zfy, Keshavarzi:2018mgv, Colangelo:2018mtw, Hoferichter:2019gzf, Davier:2019can, Keshavarzi:2019abf, Kurz:2014wya, Melnikov:2003xd, Masjuan:2017tvw, Colangelo:2017fiz, Hoferichter:2018kwz, Gerardin:2019vio, Bijnens:2019ghy, Colangelo:2019uex, Blum:2019ugy, Colangelo:2014qya} and experiment measurement, as $\Delta a_\mu\equiv a_\mu^{\text{Exp}}-a_\mu^{\text{SM}}=(25.1\pm5.9)\times 10^{-10}$. 
It is worth noting that the status of the theoretical calculation of the SM prediction has not been settled yet~\cite{Borsanyi:2020mff, Ce:2022kxy}. In this paper, we make the assumption that the apparent discrepancy is due to the contribution of new physics beyond the Standard Model.

In this work, we focus on a scenario with the contribution of  ALP as the explanation of the apparent deviation in $(g-2)_\mu$. Since the ALP is a pseudoscalar, the axion-muon coupling contributes $\Delta a_\mu$ negatively. Therefore, a model in which this is the dominant coupling can not explain the $(g-2)_\mu$ anomaly. If
the axion-photon coupling have a different sign comparing with the axion-muon coupling, the contribution can be positive~\cite{Chang:2000ii, Dedes:2001nx, Gunion:2008dg, Marciano:2016yhf, Bauer:2017ris, Bauer:2018uxu, Cornella:2019uxs, Bauer:2019gfk, Buen-Abad:2021fwq, Bauer:2021mvw}. 
Therefore, both of these couplings need to be present at the same time.
Since the axion-photon coupling comes from the interaction with hypercharge gauge field and $SU(2)_L$ gauge fields, in general, we should include both of them in addition to the muon coupling. Taking into account these considerations, we have the following effective Lagrangian at weak scale energy
\begin{align}
	& \mathcal{L}_{\rm eff}^{\text{D}\le 5}=\sum_f \frac{C_{ff}}{2}\frac{\partial^\mu a}{f_a}\bar{f}\gamma_\mu\gamma_5 f
	+\frac{g^2}{16\pi^2}C_{WW}\frac{a}{f_a}W_{\mu\nu}^i\tilde{W}^{\mu\nu, i}+\frac{g'^2}{16\pi^2}C_{BB}\frac{a}{f_a}B_{\mu\nu}\tilde{B}^{\mu\nu} ,
	\label{eq:effLag}
\end{align}
 where $W$ and $B$ are the $SU(2)_L$ and $U(1)_Y$ field strength. In this paper, we study the existing constraints on the  parameters in Eq.~\ref{eq:effLag} from collider searches,  and propose to exploit searches with final states $(\gamma \gamma) \gamma$, $(\mu^+ \mu^-) \gamma$ and $(\mu^+ \mu^-) \mu^+ \mu^-$
 at future electron-positron colliders \cite{cepc, FCC:2018evy}, with runnings in both the Tera-$Z$ and the Higgs factory modes, to search for the ALP. We found that future $Z$ factories can cover most of the parameter region of $m_a $ up to 85 GeV, which can explain the $(g-2)_\mu$ anomaly, while future Higgs factories can extend the limits to much higher masses. 

The $(g-2)_\mu$ and the constraints on the axion couplings have been extensively studied in Refs.~\cite{Bauer:2017ris, Bauer:2018uxu, Bauer:2019gfk, Buen-Abad:2021fwq, Bauer:2021mvw}. Our work extends these results in the following aspects. 

The 2-loop Barr-Zee diagram contributing to the $(g-2)_\mu$ has a nontrivial counter-term arising from the axion shift symmetry in the derivative coupling basis, which is clarified recently in Ref.~\cite{Buen-Abad:2021fwq}. Their calculations only consider the photon in the diagram. In our study, since we are interested in axion mass up to $Z$ mass, 
the contributions from the $W/Z$ gauge boson are also included.

Previous studies on the constraints from other experimental searches focus on the effect of turning on a single coupling. We start with both the fermion coupling $C_{\mu\mu}$ and the gauge boson coupling $C_{BB}$ and $C_{WW}$ due to the requirement of $(g-2)_\mu$. It opens up some unique channels, for example $Z \to (\mu^+ \mu^-)\gamma$, in the multiple parameters scenario. It differs from a previous study of $Z\to (\mu^+\mu^-)\gamma$ in Ref. [51], which only focused on $C_{\mu\mu}$ and derived $C_{\gamma Z}$ from the lepton coupling at the 1-loop level. Therefore, we conducted a reanalysis of this channel in the context of multiple parameters and particularly focused on its implications for the $(g-2)_\mu$ parameter region at future electron-positron colliders. In addition, we update the constraints from various existing experiments. For example, the muonic force study on $U(1)_{L_\mu -L_\tau}$ at CMS~\cite{CMS:2018yxg} can set limits on the axion-muon coupling, but it is missing in the previous studies. Moreover, with two gauge couplings, one of the couplings $C_{\gamma \gamma}$ or $C_{Z\gamma}$ can be very small, leading to significant changes in the axion decay branching ratio and lifetime. Therefore, some previous studies do not directly apply to this case.

We should also mention the flavor physics measurements can, in principle, offer sensitive probes to axion couplings. 
For instance, in a UV completion of the ALP Lagrangian, there may exist flavor off-diagonal derivative couplings between the ALP and leptons. Such couplings could trigger charged lepton flavor violation processes such as $\mu\to e\gamma$ and $\pi\to\mu e$ and are consequently stringently constrained by experimental results~\cite{Bauer:2019gfk,Bauer:2021mvw,Cornella:2019uxs,Dev:2017ftk}. Hence, we assume that the ALP-lepton coupling is flavor diagonal at low energy scales in the lepton mass basis. This setting implies that the UV completion of the ALP model must possess a specific flavor structure to account for the pronounced suppression of flavor off-diagonal couplings~\cite{Bauer:2021mvw,Buen-Abad:2021fwq}. Furthermore, the $C_{WW}$  coupling can induce the flavor-violating couplings in the down-quark sector through the top quark in the loop, which is severely constrained by the precision meson measurements. The $C_{WW}$ coupling can be constrained by $B^+ \to K^+ a (\mu \mu)$,
$B \to K^* a (\mu \mu)$, $B^+ \to \pi^+ a (\mu \mu)$ for $m_a < 5$ GeV~\cite{Bauer:2021mvw}. The  $C_{BB}$ coupling is less constrained since it comes in at a higher order. For $C_{\mu\mu}$ coupling, the three exotic $B$ meson decay channels provide similar constraints. There is one more channel $B_s \to \mu^+ \mu^-$, which can provide a constraint competitive with  the CMS muonic force search~\cite{CMS:2018yxg} even for large $m_a$.  At the same, such constraints depend on the flavor model, which necessarily involves more parameters. In this paper, we will assume that flavor constraints are not enough to cover the interesting parameter space. 

The paper is organized as follows. In section~\ref{sec:model}, we describe the axion low-energy model and calculate the $(g-2)_\mu$ at the 2-loop level. In section~\ref{sec:exist-and-future}, we start without the coupling between the ALP and the W boson. The signal final states are classified as $a + \gamma$ and $a + \bar{f}f$ and the existing constraints from electron-positron colliders like BaBar~\cite{BaBar:2020jma,BaBar:2016sci},Belle-II~\cite{Belle-II:2020jti} and LEP~\cite{OPAL:2002vhf,OPAL:1991acn,Jaeckel:2015jla,L3:1994shn} and the Large Hadron Collider~\cite{ATLAS:2015rsn,CMS:2018yxg,CMS:2019lwf} are discussed extensively in section \ref{sec:existConstraints}. We then discuss the constraints from exotic final states at future Tera-$Z$ and Higgs factory in section \ref{sec:futureZfactory} and extend the results including the general couplings to $W$ boson in section \ref{sec:generalGaugeCouplings}. Section \ref{sec:conclusion} contains our conclusion.

\section{The ALP properties, and contribution to $(g-2)_\mu$} 
\label{sec:model}

In this section, we present analytical results useful for our numerical study, including the  ALP couplings, relevant decay widths, and their contribution to $(g-2)_\mu$. 
Following the notation of \cite{Buen-Abad:2021fwq}, we begin with the effective Lagrangian on the basis of the fermion masses. 
\begin{align}
	\mathcal{L}_{\rm eff}^{\text{D}\le 5}
	& = \sum_f \frac{C_{ff}}{2}\frac{\partial^\mu a}{f_a}\bar{f}\gamma_\mu\gamma_5 f +\frac{\alpha C_{\gamma\gamma}}{4\pi}\frac{a}{f_a}F_{\mu\nu}\tilde{F}^{\mu\nu}+\frac{\alpha C_{\gamma Z}}{2\pi s_w c_w}\frac{a}{f_a}F_{\mu\nu}\tilde{Z}^{\mu\nu} +\frac{\alpha C_{ZZ}}{4\pi s_w^2c_w^2}\frac{a}{f_a}Z_{\mu\nu}\tilde{Z}^{\mu\nu} \nonumber \\
	&+\frac{\alpha C_{WW}}{\pi s_w^2}\frac{a}{f_a}\epsilon_{\mu\nu\rho\sigma}\partial^\mu W^\nu_+\partial^\rho W^\sigma_- + ... ~, 
\end{align}
with the coefficients
\begin{align}
	C_{\gamma\gamma}&=C_{WW}+C_{BB}, \ C_{\gamma Z}=c_{w}^2C_{WW}-s_w^2 C_{BB}, \ C_{ZZ}=c_w^4 C_{WW}+s_w^4 C_{BB},
\end{align}
where $s_w \equiv \sin \theta_{W}$,  $c_w\equiv \cos \theta_W$ with $\theta_W$ being the weak mixing angle. This EFT Lagrangian can be valid upto scale $\Lambda = 4 \pi f_a$~\cite{Buen-Abad:2021fwq,Bauer:2021mvw}. We have omitted the gluon coupling and interaction vertices containing more than three fields. In this work, we only consider $f=\mu$ for simplicity.

For illustrative purposes, we start with a simple case $C_{WW} = 0$ to present our analytic results and discuss experimental constraints before we present the full numerical results with $C_{WW} \neq 0$. This simple choice is also favored by Electroweak Precision Data (EWPD) (see, e.g., Figure 27 of Ref.~\cite{Bauer:2017ris}). In this case, we have three free parameters
\begin{align}
	\{ ~m_a, \quad C_{\mu\mu}, \quad C_{BB} ~ \}.
\end{align}
In this case, the exotic $Z$ decay  $Z \to \gamma a$ is induced by $C_{Z \gamma} = - s_w^2 C_{BB} = - s_w^2 C_{\gamma \gamma}$. 
 The partial widths of the  exotic $Z$ decay and the two axion decay channels $a \to \mu^+ \mu^-$ and $a \to \gamma \gamma$ are 
 \begin{align}
	& \Gamma(Z\to \gamma a)=\frac{\alpha^2(m_Z)m_Z^3}{96\pi^3s_w^2c_w^2f_a^2}\left|C_{\gamma Z}^{\rm{eff}}\right|^2\left(1-\frac{m_a^2}{m_Z^2}\right)^3 , \\
	&\Gamma(a\to \mu^+\mu^-)=\frac{m_am_\mu^2}{8\pi f_a^2}\left|C_{\mu\mu}^{\rm eff}\right|^2\sqrt{1-\frac{4m_\mu^2}{m_a^2}} , \\{}
	&\Gamma(a\to\gamma\gamma)=\frac{\alpha^2m_a^3}{64\pi^3f_a^2}\left|C_{\gamma\gamma}^{\rm{eff}}\right|^2,
\end{align}
where the effective couplings are
\begin{align}	
	&C_{\gamma\gamma}^{\rm{eff}}=C_{\gamma\gamma}+C_{\mu\mu}\left[1+\frac{m_{\mu}^2}{m_a^2}\cdot \mathcal{F}\left(\frac{m_a^2}{m_\mu^2}\right)\right]+\mathcal{O}(\alpha), \label{eq:C_AA_eff} \\
	&C_{\gamma Z}^{\rm{eff}}=C_{\gamma Z}+C_{\mu\mu}(\frac{1}{4}-s_w^2)\left[1+\frac{m_\mu^2}{m_a^2-m_Z^2}\cdot \left(\mathcal{F}\left(\frac{m_a^2}{m_\mu^2}\right)-\mathcal{F}\left(\frac{m_Z^2}{m_\mu^2}\right)\right)\right]+\mathcal{O}(\alpha),
	\label{eq:C_AZ_eff}  \\
	&C_{\mu\mu}^{\text{eff}}=C_{\mu\mu}+\mathcal{O}(\alpha^2), \\
	&\mathcal{F}(x)\equiv\text{ln}^2\left(\frac{\sqrt{x(x-4)}-x+2}{2}\right),
	\end{align}
with the fine structure constant evaluated at $Z$-pole $\alpha(m_Z)\approx 1/127.9$. Since the axion gauge boson coupling $C_{\gamma\gamma}, ~ C_{\gamma Z}$ enters 
to the $(g-2)_\mu$ at 1-loop, it is necessary to include the 1-loop contributions in the effective couplings. The effective couplings presented here are consistent with those in Ref.~\cite{Bauer:2017ris}.
 
\subsection{The contribution to the $(g-2)_\mu$}

\begin{figure}[htbp]
        \centering
        \includegraphics[width=0.98 \textwidth]{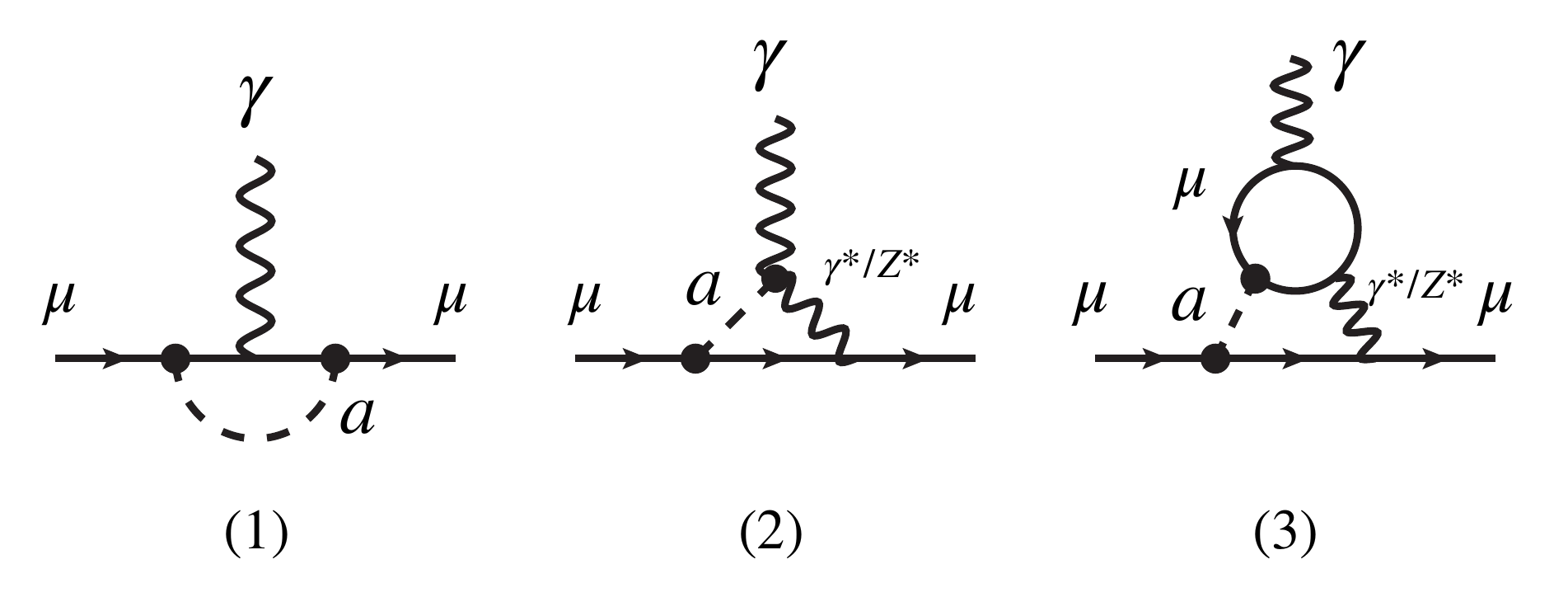}
        \caption{The loop diagrams with ALP for $(g-2)_\mu$ up to 2-loop level. The gauge bosons in diagrams (2) and (3) are $\gamma$ and $Z$ bosons.
        }
\label{fig:g-2}
\end{figure}

In this leptophilic axion setup, there are three types of diagrams, shown in Fig.~\ref{fig:g-2},  contributing to the $(g-2)_\mu$ and potentially giving an explanation to $\Delta a_\mu$ \cite{Buen-Abad:2021fwq}. The first two are 1-loop diagrams from muon and photon/$Z$ boson couplings, and the third one is the Barr-Zee type diagram from muon coupling only. The first diagram $a_\mu^{(1)}$ and third diagram $a_\mu^{(3)}$contribute negatively to $\Delta a_\mu$, with $a_\mu^{(1, 3)}\propto -C_{\mu\mu}^2$. This can be understood by the fact that the axion is CP-odd pseudoscalar, resulting in an extra minus sign in comparison with the scalar. Therefore, to explain the $(g-2)_\mu$ anomaly, the contribution of the second diagram $a_\mu^{(2)}\propto -C_{\mu\mu}C_{\gamma\gamma}$ has to be positive, implying that the couplings $C_{\mu \mu}$ and $C_{\gamma \gamma}$ should have different sign~\cite{Marciano:2016yhf,Bauer:2017ris,Chang:2000ii}. Without loss of generality, from now on, we assume that $C_{\gamma\gamma}/C_{BB}$ is positive while $C_{\mu\mu}$ is negative.

The contributions of the diagrams in Fig.~\ref{fig:g-2} are given by
\begin{align}
      \Delta a_1 &= -\frac{C_{\mu\mu}^2}{16\pi^2}\frac{m_\mu^2}{f_a^2}\left[1+2x+x(1-x)\text{ln}(x)+\frac{2x(x-3)}{x-4}\sqrt{x(x-4)}\text{ln}\left(\frac{\sqrt{x-4}+\sqrt{x}}{2}\right)\right],\label{eq:g-2-a}\\
       & ~~~~~~~~~~~~~\Delta a_{2\gamma} = -\frac{\alpha C_{\mu\mu}C_{\gamma\gamma}}{8\pi^3}\frac{m_\mu^2}{f_a^2}\cdot h_\gamma\left(x,\Lambda\right),
		\label{eq:g-2-bg} \\
   & ~~~~~~~~~~~~~\Delta a_{2Z} =\frac{\alpha C_{\gamma Z}C_{\mu\mu}\left(4s_w^2-1\right)}{32\pi^3c_w^2s_w^2}\frac{m_\mu^2}{f_a^2}\cdot h_Z(x,y,\Lambda) \label{eq:g-2-bz}, \\
	 & ~~~~~~~~~~~~~\Delta a_{3\gamma}=-\frac{\alpha Q_f^2 C_{\mu\mu}C_{ff}}{8\pi^3}\frac{m_\mu^2}{f_a^2}\left[H_\gamma+h_\gamma(x,\Lambda) \right],
		\label{eq:g-2-cg}\\
	 & ~~~~~~~~~~~~~\Delta a_{3Z}=\frac{\alpha C_{ff}C_{\mu\mu}(4s_w^2-1)Q_f(\frac{1}{2}T_3^f-s_w^2Q_f)}{32\pi^3c_w^2s_w^2}\frac{m_\mu^2}{f_a^2}\left[H_Z+h_Z(x,y,\Lambda) \right],
		\label{eq:g-2-cz} 
	\end{align}

where the $H_{\gamma/Z}$ and $h_{\gamma/Z}$ functions are 
\begin{align}
	H_Z=&\int_0^1 dz  \left\{ \frac{\Delta^2}{\Delta^2-m_a^2}\left[-\frac{m_a^2(m_a^2+2m_\mu^2)B(m_\mu^2,m_a,m_\mu)}{3(m_a^2-m_Z^2)m_\mu^2}+\frac{m_a^6}{6m_\mu^4(m_a^2-m_Z^2)}\text{ln}\left(\frac{m_a^2}{m_\mu^2}\right)\right] \right.\nonumber\\
	&\left.+\frac{\Delta^2}{\Delta^2-m_Z^2}\left[\frac{m_Z^2(2m_\mu^2+m_Z^2)B(m_\mu^2,m_Z,m_\mu)}{3m_\mu^2(m_a^2-m_Z^2)}+\frac{m_Z^6}{6m_\mu^4(m_Z^2-m_a^2)}\text{ln}\left(\frac{m_Z^2}{m_\mu^2}\right)\right]\right.\nonumber\\
	&\left.+\frac{\Delta^2}{3m_\mu^2}+\frac{\Delta^4(\Delta^2+2m_\mu^2)B(m_\mu^2,m_\mu,\Delta)}{3m_\mu^2(\Delta^2-m_a^2)(\Delta^2-m_Z^2)}+\frac{\Delta^8}{6m_\mu^4(\Delta^2-m_a^2)(\Delta^2-m_Z^2)}\text{ln}\left(\frac{m_\mu^2}{\Delta^2}\right)\right\} , 
\end{align}
\begin{align}
	H_\gamma =&\int_0^1 dz\frac{\Delta^2}{2(\Delta^2-m_a^2)}\left[-\frac{2}{3}\frac{m_a^2+2m_\mu^2}{m_\mu^2}B(m_\mu^2,m_a,m_\mu)+\frac{2}{3}\frac{\Delta^2+2m_\mu^2}{m_\mu^2}B(m_\mu^2,m_\mu,\Delta)\right. \nonumber\\
		&\left.+\frac{\Delta^4}{3m_\mu^4}\text{ln}\left(\frac{m_\mu^2}{\Delta^2}\right)+\frac{m_a^4}{3m_\mu^4}\text{ln}\left(\frac{m_a^2}{m_\mu^2}\right)+\frac{2}{3}\frac{\Delta^2-m_a^2}{m_\mu^2}   \right].\\
	h_\gamma=&\text{ln}\left(\frac{\Lambda^2}{m_\mu^2}\right)+2-\frac{x^2}{6}\text{ln}(x)+\frac{x}{3}+\frac{x+2}{3}\sqrt{x(x-4)}\text{ln}\left(\frac{\sqrt{x-4}+\sqrt{x}}{2}\right), \\
	h_Z=&\text{ln}\left(\frac{\Lambda^2}{m_Z^2}\right)+\frac{x(x+2)B(m_\mu^2,m_a,m_\mu)}{3(x-y)}+\frac{y(2+y)B(m_\mu^2,m_Z,m_\mu)}{3(y-x)}   \nonumber \\
	&+ \frac{\left(x+6+y\right)}{3}-\frac{6x+y^3-6y}{6(x-y)}\text{ln}\left(\frac{x}{y}\right)-\frac{x^2+xy-6+y^2}{6}\text{ln}\left(x\right), \label{eq:countertermz}
\end{align}
with
\begin{align}
	& x \equiv\frac{m_a^2}{m_\mu^2} \  ,\  y\equiv\frac{m_Z^2}{m_\mu^2} \ ,   \ \Delta^2\equiv \frac{m_f^2}{z(1-z)}, \\
	&B(m_\mu^2,m_{a/Z},m_\mu)=\sqrt{m_{a/Z}^2(m_{a/Z}^2-4m_\mu^2)}\ln\left(\frac{\sqrt{m_{a/Z}^2-4m_\mu^2}+m_{a/Z}}{2m_\mu}\right)/m_\mu^2,\\
	&B(m_\mu^2,m_{\mu},\Delta)=\sqrt{\Delta^2(\Delta^2-4m_\mu^2)}\ln\left(\frac{\sqrt{\Delta^2-4m_\mu^2}+\Delta}{2m_\mu}\right)/m_\mu^2.
\end{align}
The $B$ function is the \textbf{DiscB} function in \texttt{Mathematica} \texttt{Package X}~\cite{Package}. $\Lambda$ is the loop calculation UV cutoff scale. In our muonphilic scenario, the loop fermion is restricted to the muon lepton only, i.e., $f=\mu$. Explicit calculations have been performed in Refs.~\cite{Marciano:2016yhf, Bauer:2017ris, Buen-Abad:2021fwq}, with the inner loop of the Barr-Zee diagram using off-shell axion and photon propagators when connecting to the muon lines \cite{Buen-Abad:2021fwq}. The effective $a F \tilde F$ vertex function is then inserted into the loop, which leads to the final answer of $(g-2)_\mu$. As an appropriate approximation, we only keep effective vertex function result that is linear in the on-shell photon's momentum. 

In addition, we add the contribution with internal $Z$ boson for completeness, denoted as $\Delta a_{2Z}$ and $\Delta a_{3Z}$, especially because we are interested in large $m_a$ comparable to $m_Z$. Moreover, for non-zero $C_{WW}$, it is possible that $C_{\gamma Z}$ can be much larger than $C_{\gamma\gamma}$. Thus the diagram with $Z$ boson in the loop is no longer negligible. The result in the second diagram of Fig.~\ref{fig:g-2} with an inner $Z$ boson takes the form $\Delta a_{2Z}\propto \text{log}\frac{\Lambda^2}{m_Z^2}+3/2$ when expanded at large $y$, which is consistent with calculations of Ref.~\cite{Bauer:2017ris}. In the third diagram of Fig.~\ref{fig:g-2},  the contribution with internal $Z$ boson to the Barr-Zee diagram is not small due to the counter-term (explained in detail below). 
Lastly, due to the ALP's antisymmetric coupling to the $W^\pm$ gauge bosons, the contribution of the 2-loop Barr-Zee diagram with the W boson in the loop is zero~\cite{Bauer:2021mvw,Ilisie:2015tra}.

The calculations in Eq.~(\ref{eq:g-2-a}--\ref{eq:g-2-cz}) are done with shift invariant derivative coupling  $\partial^\mu a \bar{f}\gamma_\mu \gamma_5 f$, which has a subtlety in the last diagram. A direct 2-loop calculation of the third diagram leads to the $H$ function. However, the vertex function of the inner loop is ambiguous since it is linearly divergent. 
For the inner loop leading to effective $a F \tilde F$ vertex, it has to vanish when sending the fermion mass in the inner loop to infinity to preserve the shift symmetry of ALP. This fact is not satisfied for $H$ function alone. Therefore, one has to add the counter term function $h$ to fix this problem as shown in Ref.~\cite{Buen-Abad:2021fwq}. The counter term here works as shift symmetry restoration in the derivative coupling basis and should not be mixed with the counter terms in renormalization procedure.

It turns out that the function $h$ dominates over $H$ in the parameter space we are interested in. 
It is suggestive to compare with the results in the other operator basis. One can do a chiral rotation to eliminate the derivative term, which is equivalent to doing the following substitution,
\begin{align}
	\frac{C_{ff}}{2}\frac{\partial^\mu a}{f_a}\bar{f}\gamma_\mu\gamma_5f &\to -\frac{C_{ff}m_f}{f_a}a\bar{f}i\gamma_5f +\frac{g'^2(Y_L^2+Y_R^2) C_{ff}}{32\pi^2 f_a}a B^{\mu\nu}\tilde{B}_{\mu\nu} + \frac{g^2 T_3^2 C_{ff}}{32\pi^2 f_a}a W_{3\mu\nu}\tilde{W_3}^{\mu\nu}\nonumber\\
	&+\frac{g g' T_3 Y_L C_{ff}}{16\pi^2 f_a}a B_{\mu\nu}\tilde{W}_3^{\mu\nu} + \cdots\\
	&\to -\frac{C_{ff}m_f}{f_a}a\bar{f}i\gamma_5f+\frac{\alpha Q^2 C_{ff}}{4\pi f_a}a F_{\mu\nu}\tilde{F}^{\mu\nu} + \frac{\alpha C_{ff}Q(\frac{1}{2}T_3-s_w^2Q)}{2\pi c_w s_w f_a} aF^{\mu\nu}\tilde{Z}_{\mu\nu} \nonumber\\
	&+ \frac{\alpha C_{ff}(s_w^4Q^2-s_w^2QT_3+\frac{T_3^2}{2})}{4\pi c_w^2 s_w^2 f_a}a Z^{\mu\nu}\tilde{Z}_{\mu\nu} + \cdots
	\label{eq:operator-change}
\end{align}
where $Y_L,~Y_R,~T_3$ are the hypercharge of left-handed, right-handed, and weak isospin of fermions, respectively. We have omitted the terms which are higher order in $f_a^{-1}$. The anomalous $aF^{\mu\nu}\tilde{Z}_{\mu\nu}$ part can be determined by the calculation of anomalous triangle diagrams with massless fermions running in the loop. Because of the Furry theorem, only the vector part of $Z$ coupling to fermions contributes to the diagrams. 
When using the $i a \bar{\mu }\gamma_5 \mu$ operator, the result for the first diagram is not changed, while the third diagram returns exactly the $H$ function. Since there is no shift symmetry, the results are exact. Furthermore, the extra $a F^{\mu\nu} \tilde F_{\mu\nu}$ coupling in Eq.~(\ref{eq:operator-change}) leads to a new contribution in the second diagram, which matches the counter-term contribution as the $h$ function. We have done the calculation using Package-X~\cite{Package}. The results of the $H$ functions are consistent with other calculations. For the 2-loop calculations, when the loop fermion mass $m_f$ and the ALP mass $m_a$ satisfy the condition ($m_\mu \ll m_a, m_f$), our results of the $H$ functions coincide with the 2-loop results in Ref.~\cite{Chang:2000ii} calculated with $i\bar \mu \gamma_5 \mu$. 

\begin{figure}[h!]
    \centering
    \includegraphics[width=0.6 \textwidth]{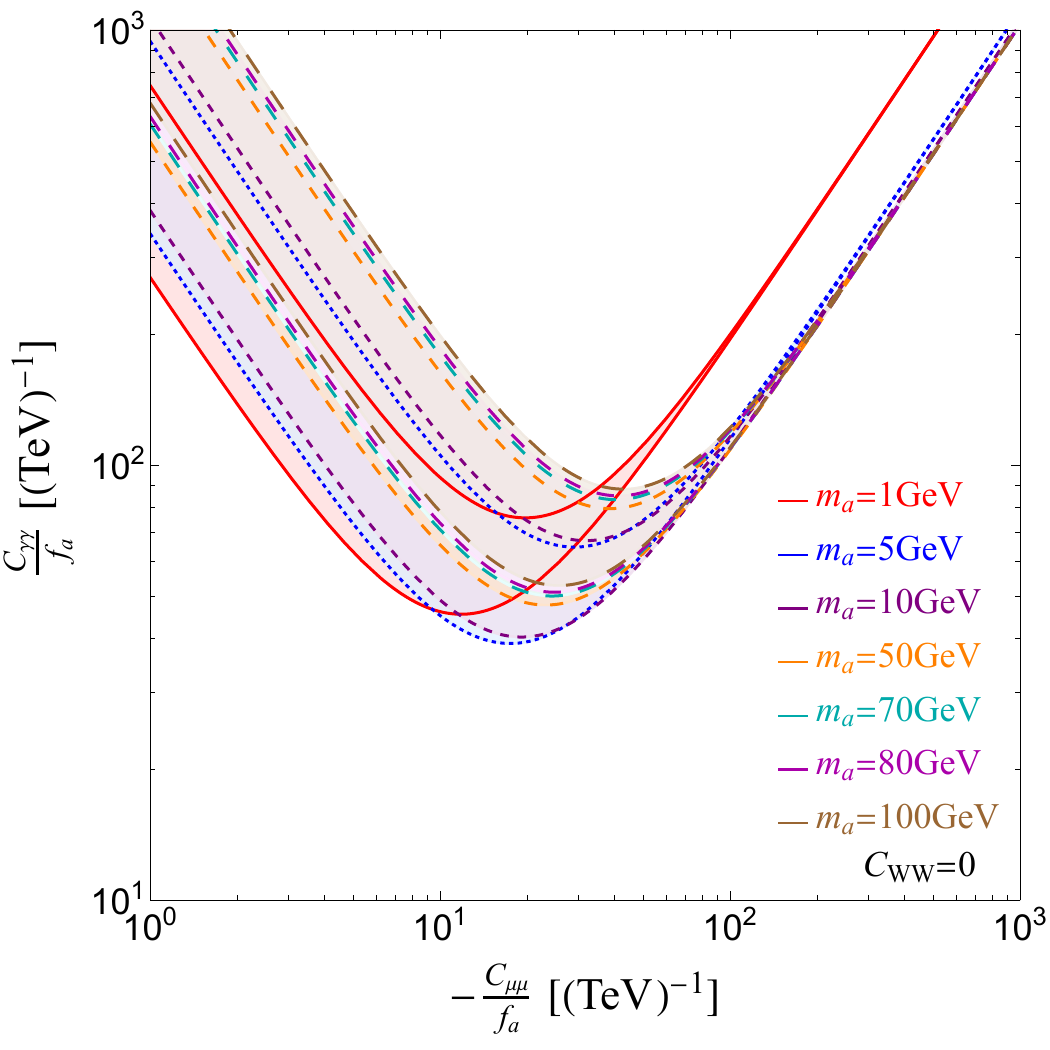}
	\caption{The bands in the axion couplings which can give an explanation to the $(g-2)_\mu$ anomaly at $2\sigma$ confidence level. Different bands correspond to different choices of axion mass $m_a$.
		We set the cut-off scale, $\Lambda$, to be 1 TeV and the axion-W coupling $C_{WW}=0$.
		}
	\label{fig:parameter-g-2}
    \end{figure}

In Fig.~\ref{fig:parameter-g-2}, using Eq.~(\ref{eq:g-2-a}--\ref{eq:g-2-cz}), we show the parameter space which can give an explanation to the $(g-2)_\mu$ anomaly at $2\sigma$ confidence level.
The axion mass is chosen to be $1,~5, ~10, ~50, ~70, ~80,  \text{and} ~100$ GeV respectively, which will also serve as benchmark cases for the rest of the paper.   In the shaded regions between different color lines,  an explanation of the $(g-2)_\mu$ anomaly is possible. In these regions, the decay to di-muon dominates over the di-photon except at high mass regions around $Z$-pole. For the upper-right region, the $C_{\mu \mu}$ coupling is large and leads to a large negative contribution, which requires a large $C_{\gamma \gamma}$ coupling to cancel it. For larger couplings, this cancellation needs to be more precise. Hence, there is less flexibility for $C_{\gamma \gamma}$, and the band on the upper-right region is much thinner than the upper-left region.

The results presented in Figure 2 indicate that a minimum value of $\frac{C_{\gamma\gamma}}{f_a}$ can be around 40 ${\rm TeV}^{-1}$ to explain the $(g-2)_\mu$ anomaly. This implies that the EFT Lagrangian given in Eq. 2 should be valid up to the cutoff $\Lambda = 4\pi f_a \sim 300$ GeV, which in turn suggests that $f_a \ll v_{\rm SM}$, with $v_{\rm SM} = 246$ GeV being the SM electroweak breaking scale. Consequently, the applicability of the EFT to the underlying UV model is a challenge, and the constraints on the UV complete model may be important in phenomenological constraints on EFT. These challenges have been comprehensively discussed in Ref.~\cite{Buen-Abad:2021fwq}, which notes that the flavor off-diagonal derivative coupling of ALP is inevitable, and therefore lepton flavor violating processes should be considered. We take all relevant points into account, as recommended in Ref.~\cite{Buen-Abad:2021fwq}. Our findings suggest that there is still room for ALP $(g-2)_\mu$ explanation, which could be explored in future collider searches.

\section{The constraints from existing searches and projection of future probes}
\label{sec:exist-and-future}

The model of ALP with an explanation for the $(g-2)_\mu$ anomaly can lead to a rich set of experimental signals.
For example, there are many existing experiments searching for light new particles in a similar mass range, which can set stringent limits on ALP couplings. In section~\ref{sec:existConstraints}, we will go through the existing experiments and check how they can constrain the above parameter space. It turns out that most of the interesting parameter space capable of explaining the $(g-2)_\mu$ anomaly is still viable under the existing constraints.

\begin{figure}[htbp]
	\centering
	\includegraphics[width=0.65 \textwidth]{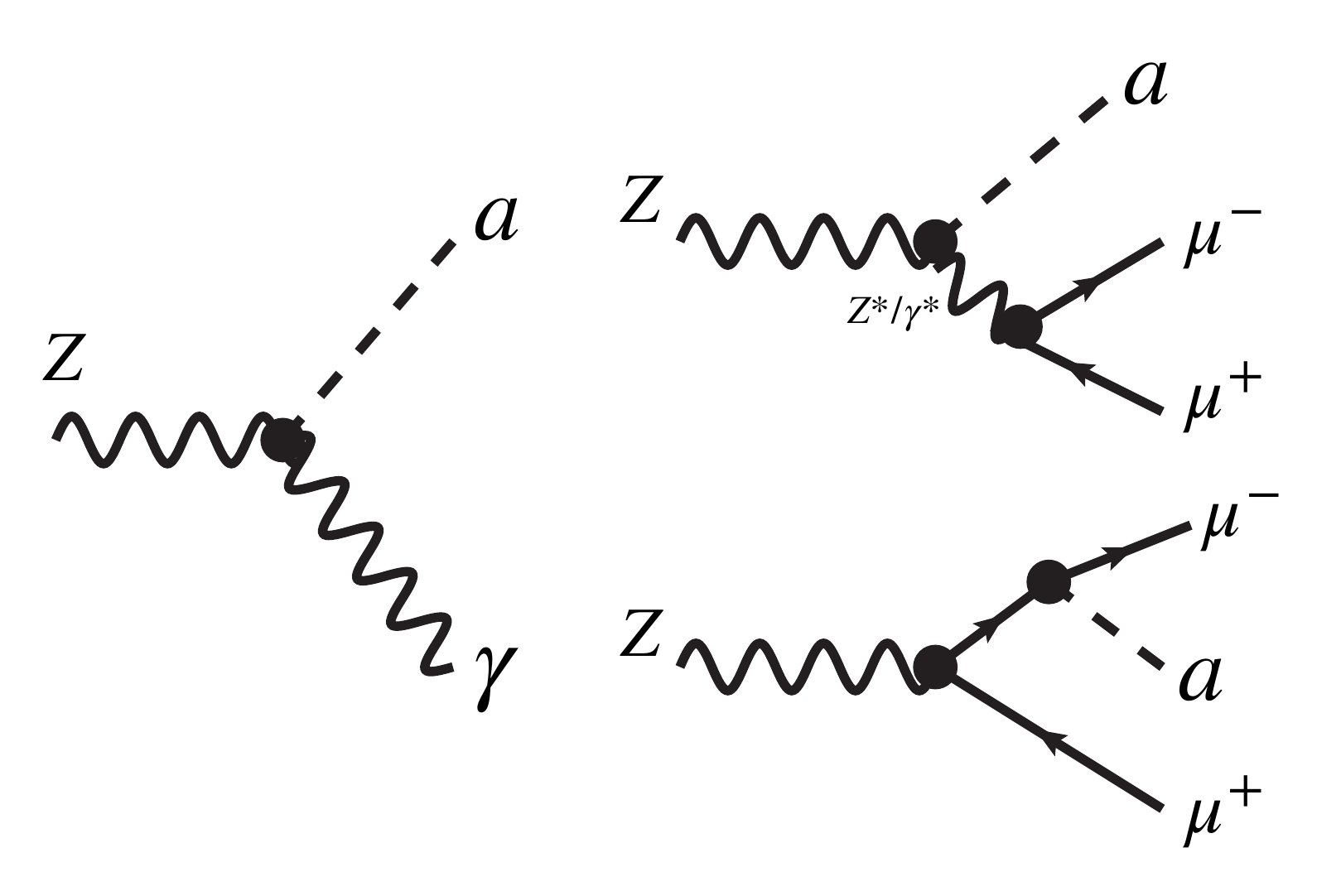}
	\caption{The Feynman diagrams for exotic $Z$ decay channels $Z\to a \gamma$ and $ Z\to a \mu^+\mu^-$ respectively.}
	\label{fig:z-diagrams}
\end{figure}

\begin{figure}[htbp]	
	\includegraphics[width=0.47\linewidth]{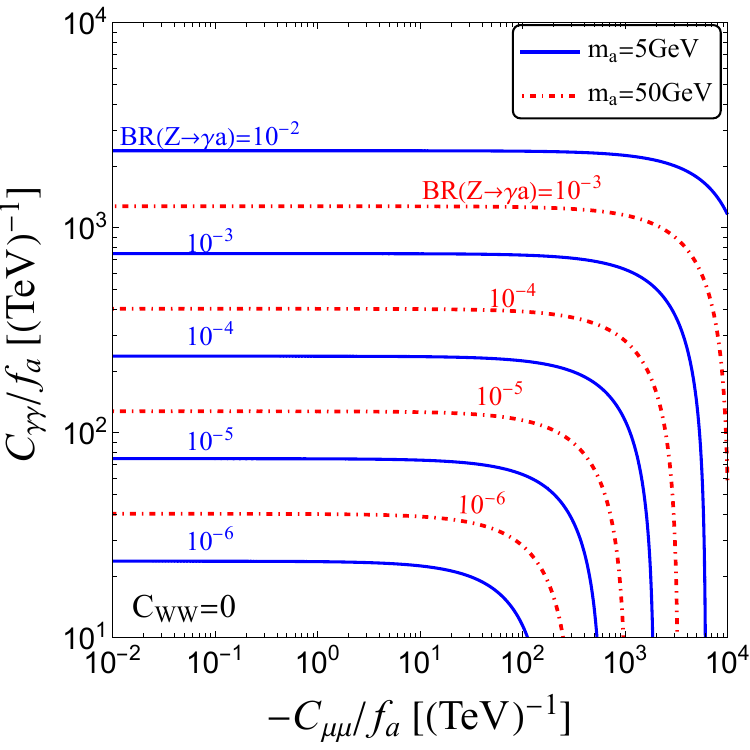}
	\includegraphics[width=0.47\linewidth]{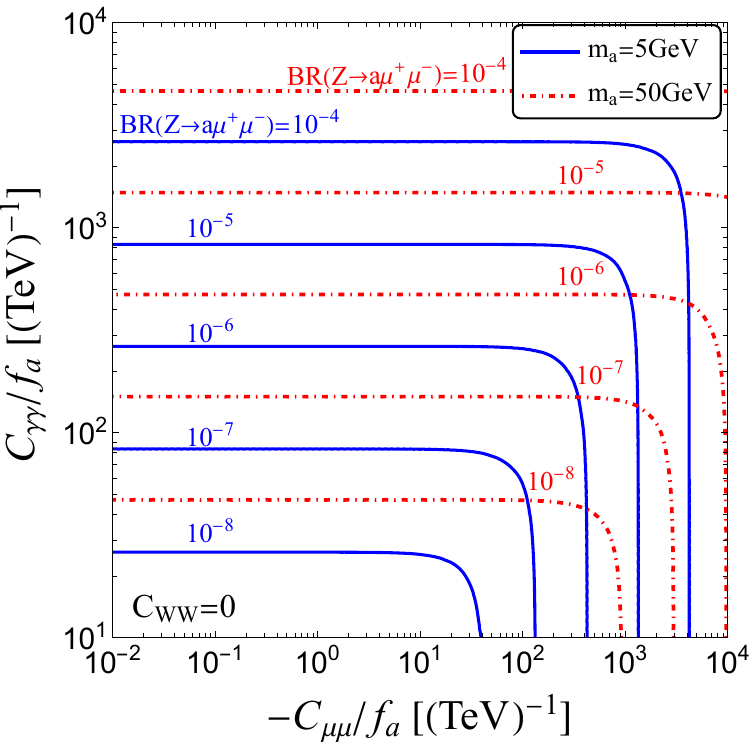}
	\caption{The exotic $Z$ decay branching ratios for $Z\to a \gamma$ (left panel) and $ Z\to a \mu^+\mu^-$ (right panel) respectively. As examples, we choose $C_{WW}=0,~m_a=5 \ \text{and} \ 50\ \text{GeV}$.
	}
	\label{fig:z-decayBR}
\end{figure}

In addition, the above couplings can lead to exotic $Z$ decay $Z \to a\gamma$ and $Z \to a \mu^+ \mu^-$, as shown in Fig.~\ref{fig:z-diagrams}, with relevant branching ratios presented in Fig.~\ref{fig:z-decayBR}. These are the main decay channels we will be considering in this paper. 
We discuss the limits from the $Z$-pole run at future electron-positron colliders in section~\ref{sec:futureZfactory} and found it can decisively exclude the ALP solution up to $m_a \sim 85$ GeV.
In this section, we will focus on the simpler case of  $C_{WW}=0$. This gives a simple picture of physics. In section~\ref{sec:generalGaugeCouplings}, we will present the numerical results with non-zero $C_{WW}$ and discuss the difference with the simpler case. 

\subsection{Constraints from current results of light particle searches}
\label{sec:existConstraints}

We focus on two final states, one is $a+ \gamma$ and the other is $a+ \bar{f}f$. In addition, axion decay channels,  $a \to \gamma \gamma$ and $a \to \mu^+ \mu^-$, are considered. In the following, we will go through the relevant experiments and set the limits on the ALP couplings to muon and photon.

\subsubsection{Constraints from searches for final states of $a + \gamma$}
\label{subsec:contraintsonagamma}
\begin{figure}[htbp]
    \centering
    \includegraphics[width=1 \textwidth]{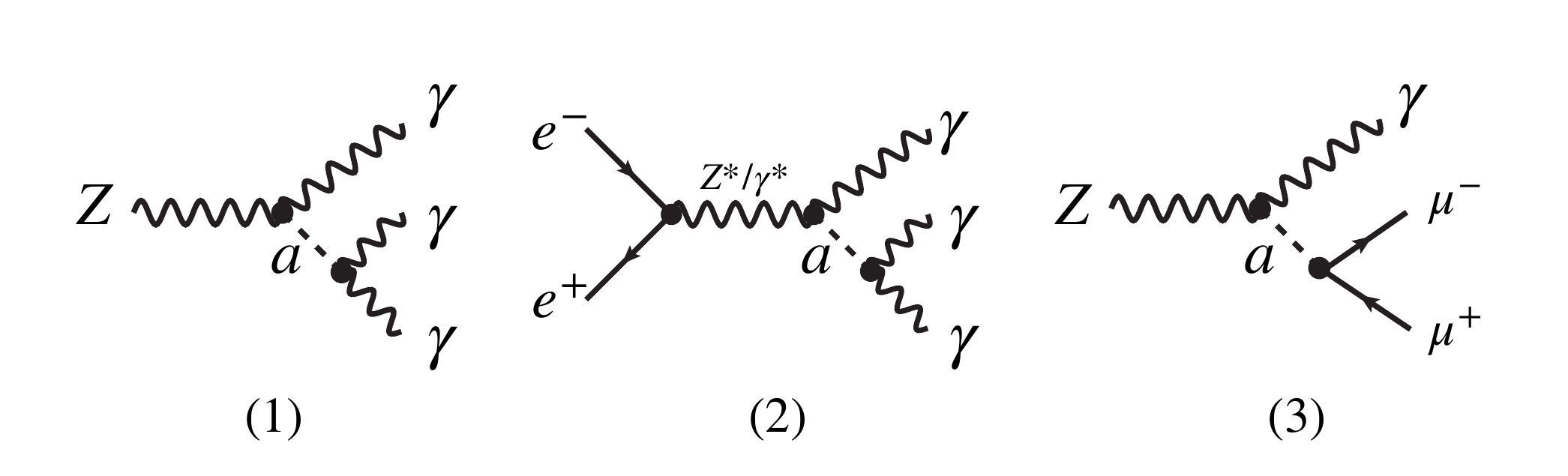}
    \caption{Feynman diagrams corresponding to $a+\gamma$ final states in various experimental searches. The diagram (1) is related to LEP/L3/ATLAS~\cite{Jaeckel:2015jla,L3:1994shn,ATLAS:2015rsn} with on-shell $Z\to (\gamma\gamma)\gamma$, while diagrams (2) and (3) are related to OPAL searches for $e^+e^-\to (\gamma\gamma)\gamma$ via off-shell $\gamma/Z$~\cite{OPAL:2002vhf} and on-shell $Z\to (\mu^+\mu^-)\gamma$~\cite{OPAL:1991acn} respectively.}
    \label{fig:agamma-finalstates}
\end{figure}

ALP together with a photon can show up as final states from either exotic $Z$ decay $Z \to a \gamma$ or the s-channel off-shell photons and $Z$ bosons production $e^+ e^- \to \gamma^* / Z^* \to a \gamma$, though couplings $C_{\gamma\gamma} $ and $ C_{\gamma Z}$ (for $C_{WW}=0$, $C_{\gamma Z}$ is fixed by $C_{\gamma\gamma}$) as shown in Fig.~\ref{fig:agamma-finalstates}. Since both $C_{\mu \mu} $ and $C_{
\gamma \gamma} $ are non-zero in order to account for the $(g-2)_\mu$ anomaly, the ALP will  decay to $\mu^+ \mu^-$ and $\gamma \gamma$. Therefore, the experimental searches for $(\gamma \gamma)+ \gamma$ and $(\mu^+\mu^-) +\gamma$ final states, where the bracket indicates the two particles inside form a resonance, should be sensitive to this class of models. 
\begin {itemize}
	\item \textit{Two photon final state $\gamma\gamma$}. There have been a large number of relevant searches at LEP and LHC which focus on multi-photon final state. For very low mass ALPs, the two photons decayed from boosted ALPs are too collimated to be resolved by the detector. Therefore, the two photons will be recognized as one single photon. As a result, in the low mass region the $2\gamma$ final state search is more relevant. 
	LEP-I inclusive di-photon searches $ e^+ e^- \to  X + 2\gamma$ is exploited to cover the low mass ALPs in Ref.~\cite{Jaeckel:2015jla, Liu:2017zdh}. The $Z$-pole production $Z \to a \gamma$ and virtual photon production $e^+ e^- \to \gamma^* / Z^* \to a \gamma$ have been considered and the limits on $C_{\gamma Z}$ and $C_{\gamma \gamma}$ have been derived with the assumption ${\rm BR}(a \to \gamma\gamma ) =100\%$. With the inclusion of $a\to \mu^+ \mu^-$, it  can still place the most stringent bound for $m_a = 1$ GeV as shown in Fig.~\ref{fig:ma-varying-final-plot} and for $m_a <4 $ GeV  as shown in the right panel of Fig.~\ref{fig:finalresults}.     
	
	\item \textit{Three photon final state $(\gamma \gamma) \gamma$ where the pair $(\gamma\gamma)$ come from the ALP decay\footnote{In this paper, we use the convention that two particles inside a bracket come from the ALP.}}.
	The searches based on this final state are relevant for higher ALP masses. L3 and ATLAS have looked for exotic $Z$ decay $Z\to 3\gamma$~\cite{ATLAS:2015rsn,L3:1994shn}, while OPAL collaboration has searched for off-shell $\gamma /Z$ process $e^+e^-\to X(\gamma\gamma)\gamma$~\cite{OPAL:2002vhf}. The ATLAS experiment sets a limit $\text{BR}(Z\to \gamma\gamma\gamma)<2.2\times 10^{-6}$ \cite{ATLAS:2015rsn}, which can be used to put constraints on $M_a \lesssim \ 70$ GeV. There is no limit on $m_a > 70$ GeV due to the experimental requirement $p_T^\gamma > 17\ \text{GeV}$~\cite{Knapen:2016moh}. Recently, Belle-II has searched for the channel $e^+e^- \to \gamma a \to 3\gamma $ at electron-positron colliders for ALP mass range $0.2<m_a<9.7 \ \text{GeV}$~\cite{Belle-II:2020jti}, assuming $\text{BR}(a\to \gamma\gamma) = 100\%$. After including the ALP decay branching ratio to muons, these experiment results can be recast to set stringent limits on axion-photon-photon coupling $C_{\gamma\gamma}$.  
In the right panel of Fig.~\ref{fig:finalresults}, we have assumed that for a given $C_{\gamma\gamma} \ \text{and} \ m_a$, the coupling to muon $C_{\mu\mu}$ is the minimal value which can give an explanation to the $(g-2)_\mu$ anomaly. In this case, one can see the limits on $C_{\gamma\gamma}/f_a$ should be $\lesssim 300{\rm \ TeV}^{-1}$ by LEP-I ($2\gamma$) for $m_a < 10$ GeV, and $\lesssim 100{\rm \  TeV}^{-1}$ by L3 $3\gamma$ for $4 \ {\rm GeV}< m_a < m_Z$. The ATLAS ($Z \to 3\gamma$) search is a little stronger than L3 ($Z \to 3\gamma$), but is narrower in the mass range covered. OPAL ($3\gamma$) can set limits on $C_{\gamma\gamma}/f_a < \mathcal{O}(150)~ {\rm TeV}^{-1}$ for $ m_a > 20~{\rm GeV} $  and can extend to limit to $m_a \lesssim 190~{\rm GeV}$~\cite{Liu:2017zdh}.

	\item \textit{Muon pair with a radiated photon $(\mu^+\mu^-)+\gamma$}. If ALP decays to muons, the experimental search for final state $\mu^+\mu^- \gamma $ is relevant. The OPAL collaboration has studied  pair of leptons $\ell^+ \ell^-$ plus a radiated photon from $Z$ decay~\cite{OPAL:1991acn}, where $\ell = e,~\mu,~\tau$, and set limits on exotic $Z$ decay $\text{Br}(Z\to\mu^+\mu^-\gamma)<5.6\times 10^{-4}$. They have also set limits for new resonance from $Z \to X \gamma$ with $X \to \ell^+ \ell^-$ for $60~{\rm GeV}< m_X < 84$ GeV. It has been used to constrain $C_{Z\gamma}$ while assuming axion decay $\text{Br}( a \to \ell^+\ell^- )= 100~(10)\%$ in Ref.~\cite{Bauer:2017ris}. We adapt the results to our model with both $C_{\gamma \gamma}$ and $C_{\mu\mu}$ couplings.
	However, with the choice $C_{WW}=0$,  it only shows up in the left panel in Fig.~\ref{fig:finalresults}, where our choice of minimal $C_{\gamma \gamma}$ leads to larger ${\rm BR}(a \to \mu^+ \mu^-)$.
	In the right panel of Fig.~\ref{fig:finalresults}, this limit is not relevant and the constraints are dominated by the photon final state searches. 
\end{itemize}

\subsubsection{Constraints from searches for final states of  $a + \bar{f}f$}
\label{sec:affbar}
\begin{figure}[htbp]
    \centering
    \includegraphics[width=1 \textwidth]{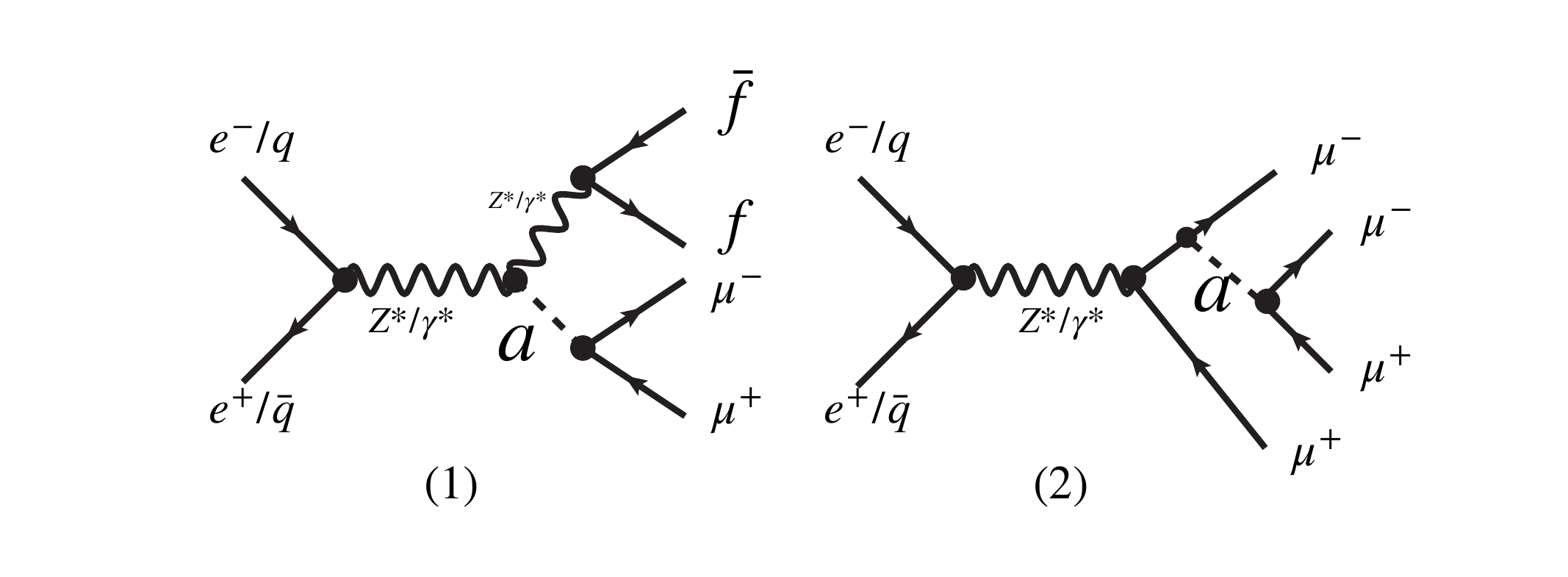}
    \caption{Feynman diagrams corresponding to $a+f\bar{f}$ final states experiments searches. Diagram (1) are related to collider searches $e^+e^-/pp\to (\mu^+\mu^-)f\bar{f}$ at BaBar~\cite{BaBar:2020jma,BaBar:2016sci},CMS~\cite{CMS:2019lwf,CMS:2018yxg}, and diagram (2) are also related to  collider searches $e^+e^-/pp\to (\mu^+\mu^-)\mu^+\mu^-$ at CMS~\cite{CMS:2018yxg},BaBar~\cite{BaBar:2016sci}. }
     \label{fig:affbar}
\end{figure}

The relevant processes with this class of final states are shown in Fig.~\ref{fig:affbar}. Depending on whether the fermion is muon lepton or not, this final state can be classified into two categories.
\begin{itemize}
	\item \textit{ A muon pair together with a pair of other fermions $(\mu^+\mu^-)+\bar{f} f$}. If the associated fermions are not muons, the ALP should be generated through axion-gauge couplings alone as shown in the left panel of Fig.~\ref{fig:affbar}. In this case, to recast the experiment limits to our model, we need to rescale the constraint on axion-gauge couplings by taking into account the $a \to \mu^+ \mu^-$ branching ratio. 
	
	Several experimental searches belong to this category. The CMS collaboration has analyzed multilepton final states in search for new scalar or pseudoscalar particles, which decay to di-muon or di-electron, assuming the particle is produced in $pp$ collisions associated with top quark pairs, i.e., $pp\to t\bar{t}\phi$ with $\phi \to \mu^+ \mu^-$~\cite{CMS:2019lwf}. Recently, the BaBar collaboration~\cite{BaBar:2020jma} has published a search result for dark leptophilic scalar in the channel $e^+e^-\to \tau^+\tau^- \phi_L, ~ \phi_L\to \ell^+\ell^-$,  setting limits to leptonic couplings for $\phi_L$ in the mass range $0.04<m_{\phi_L}<7.0~\text{GeV}$. The ALP in our model can play the role for $\phi/\phi_L$ here, generated in associated with top or tau lepton pairs and subsequently decays to a pair of muons. Their results can be recast to constrain couplings $C_{\gamma\gamma}/C_{\gamma Z}$. 
	The CMS and the BaBar experiments can constrain the $|C_{\mu\mu}/f_a|\lesssim 5000~ \text{TeV}^{-1}, ~1000~\text{TeV}^{-1}$ respectively, as shown in the left panel of Fig.~\ref{fig:finalresults}.
	For the parameter region capable of explaining the $(g-2)_\mu$ anomaly, the constraints from the $(\mu^+\mu^-)+\bar{f} f$ final state searches are less stringent compared with $4\mu$ final states search.
	
	\item \textit{Four muons final state $(\mu^+\mu^-)+\mu^+\mu^-$}.    If the fermions associated with an ALP are muons, the ALPs are generated through both the coupling $C_{\gamma\gamma}$, $C_{\gamma Z}$ and $ C_{\mu\mu}$. We can not simply rescale the existing ALP coupling limit (usually assuming only one ALP coupling) since more Feynman diagrams are relevant, which could alter the cut efficiency for the signal model. Therefore, we use MadGraph5@NLO~\cite{Alwall:2014hca} and FeynRules~\cite{Alloul:2013bka} to simulate and recast the experimental bounds.
	
	There are a number of experimental searches for the $4\mu$ final states, often targeting some mediator of the so-called muonic dark force, such as the $Z'_{L_\mu-L_\tau}$ gauge boson. The BaBar collaboration~\cite{BaBar:2016sci} has carried out the $e^+e^- \to 4\mu$ channel analysis using 514 fb$^{-1}$ data, looking for exotic gauge boson $Z'$ in mass range of $0.212$--$10$ GeV. The CMS collaboration~\cite{CMS:2018yxg} has performed a similar search for $Z'$ in mass range $10$--$70$ GeV at LHC. In these results, the production and decay of the $Z'$s are both assumed to be governed by a single coupling to muon. To recast their limits, both the couplings of $C_{\mu\mu}$ and $C_{\gamma\gamma/\gamma Z}$ need to be considered due to the differences in relevant Feynman diagrams and the effect of ALP decay branching ratio. In addition, the different cut efficiencies between vector gauge bosons and pseudoscalar have been taken into account by simulation. 
	In the left panel of Fig.~\ref{fig:finalresults}, 
	the $4\mu$ final state search places the most stringent constraints on parameter spaces relevant for explaining the $(g-2)_{\mu}$ anomaly. It excludes $|C_{\mu\mu}/f_a| \ge \mathcal{O}(10)\text{TeV}^{-1}$. 
\end{itemize}

\subsubsection{Constraints from measurements of Light-by-Light scattering}
\begin{figure}[htbp]
	\includegraphics[width=0.60\linewidth]{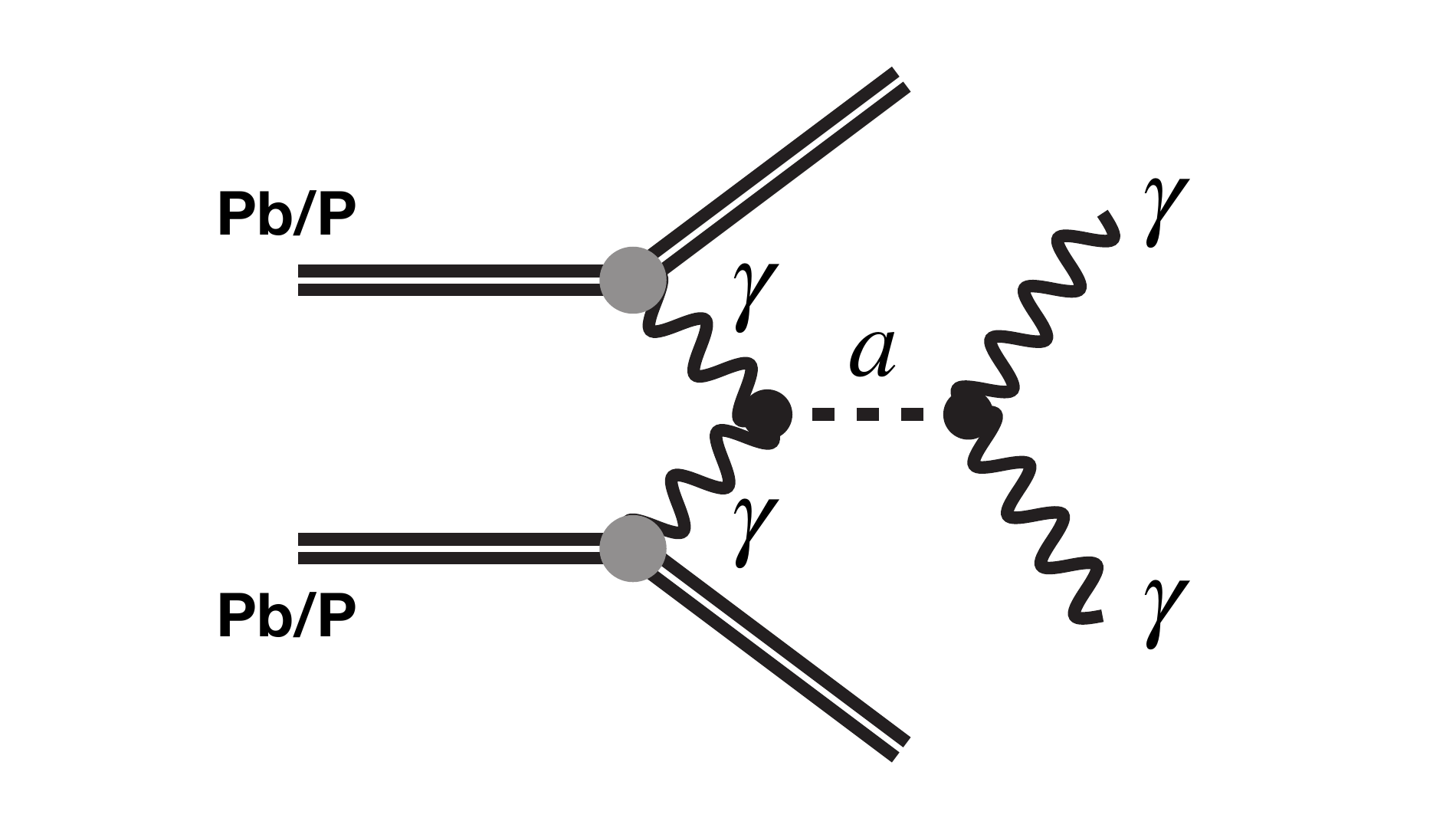}
    \caption{The photon fusion process $\gamma \gamma \to a\to \gamma \gamma $ from PbPb ion collision and pp collision at the LHC. }
	\label{fig:lbyl}
\end{figure}
Light-by-Light scattering $\gamma\gamma \to \gamma\gamma$ can occur in heavy-ion collisions and proton-proton collisions. This type of scattering can be used to probe ALPs through channel $\gamma \gamma\to a\to\gamma\gamma$. There are several experimental searches for ALPs using the same final state as the Light-by-Light scattering process. CMS~\cite{CMS:2018erd} and ATLAS~\cite{ATLAS:2020hii} have analyzed the data collected from ultraperipheral PbPb collisions at $\sqrt{s_{NN}}=5.02 ~\text{TeV}$ and set upper limits for $C_{\gamma\gamma}/f_a$ assuming $\text{BR}(a\to\gamma\gamma)=100\%$. Similarly LHC has carried out several searches for exclusive di-photon events in proton-proton collisions~\cite{CMS:2021plq,CMS:2012cve} and set limits for higher mass of ALPs. These experiments and their limitations of ALPs have been extensively studied in Ref.~\cite{Agrawal:2021dbo} and we rescale the constraints by the ALP decay branching ratio.
We have placed these limits in the Fig.~\ref{fig:finalresults} and are denoted as ${\rm PbPb}(\gamma\gamma)$ and ${\rm LHC}\gamma\gamma{\rm F}$ for PbPb and pp collisions, respectively.

\subsection{The search at future $Z$ and Higgs factories}
\label{sec:futureZfactory}

There have been several proposals for future circular electron-positron colliders, including the CEPC~\cite{CEPC-SPPCStudyGroup:2015csa,cepc} and  the FCC-ee~\cite{FCC:2018evy,Beacham:2019nyx}. 
As part of their proposed run plan, 
there is a stage running at the $Z$-pole with a target of resonant producing more than $10^{12}$ $Z$ gauge bosons, as well as runs in the Higgs factory mode \cite{cepc, Blondel:2018mad}. These are called Tera-$Z$ and Higgs factories. In particular, FCC-ee(CEPC) is proposed to run an integrated luminosity at $Z$ pole for $150~\text{ab}^{-1}(60~\text{ab}^{-1})$ and at $\sqrt{s}=240~\text{GeV}$ for $5~\text{ab}^{-1}(12~\text{ab}^{-1})$ respectively~\cite{Gao:2022lew,Bernardi:2022hny}. With numerous $Z$ and Higgs bosons and the clean environment of an electron-positron collider, it is an ideal place to look for the exotic $Z$ and Higgs decay and place limits for beyond the Standard Model physics \cite{Liu:2017lpo, Liu:2017zdh, Jin:2017guz, Chang:2018pjp, Wang:2019orr, Cheng:2019yai, Cacciapaglia:2021agf}.

The couplings $C_{\gamma Z}$ and $C_{\mu \mu}$ are crucial for an ALP  explanation of the $(g-2)_\mu$ anomaly. At the same time, they predict rare Z decays $Z \to a \gamma $ and $Z \to a \mu^+ \mu^-$, which the Tera-$Z$ factory is well equipped to test.  
The axion decays dominantly via $a \to \mu^+ \mu^-$ in the parameter space which can explain the $(g-2)_\mu$ anomaly, except when $m_a$ is close to the $Z$-pole. 
In our numerical study, we have included the 1-loop correction from muon coupling $C_{\mu \mu}$ to the ALP couplings $C_{\gamma Z/\gamma\gamma}$, shown in Eqs.~(\ref{eq:C_AA_eff}) and (\ref{eq:C_AZ_eff}). 
Since we have $C_{\gamma\gamma}^{\text{eff}}\approx C_{\gamma\gamma}+C_{\mu\mu}$, this modifies ALP decay branching ratio and the experiment constraints. 
We consider both of the ALP decays $a \to \mu^+ \mu^-$ and $a \to \gamma\gamma$, which lead to the exotic $Z$ decay final states as
$(\mu^+\mu^-)\gamma$, $(\gamma \gamma)\gamma$, and $(\mu^+\mu^-)\mu^+\mu^-$. 
The relevant SM backgrounds are simulated using MadGraph5@NLO~\cite{Alwall:2014hca}. There is one more exotic decay $(\gamma \gamma)\mu^+ \mu^-$ which is less covered by the experiments, we leave its phenomenology study to future work.
The branching ratios of relevant decay modes $Z\to a \gamma$ and $ Z\to a \mu^+\mu^-$ are given in Fig.~\ref{fig:z-decayBR} for $C_{WW}=0$, $m_a=5 ~ (50)$ GeV respectively.  
The exotic $Z$ decay searches lead to strong constraints for $m_a < m_Z$, but the limits fade away when $m_a$ close to $m_Z$. Therefore, we extend them to searches at Higgs factories with $\sqrt{s} = 240$ GeV,  based on s-channel production through off-shell $Z/\gamma$, to cover $m_a > m_Z$.

To get a more realistic estimate, we impose some basic cuts \cite{Chen:2017yel} on the kinematics of the muons and photons in the final state to suppress the SM background. 
The opposite sign muons are required to be spatially separated, $\Delta R > 0.1$.
For $a\to \mu^+\mu^-$ decay, to take into account  the resolution for the invariant mass of di-muon, we require $|m_{\mu\mu}/m_a -1|< 0.19\%$.
For $a\to \gamma \gamma$ decay, we require $|m_{\gamma \gamma}-m_a| < 1$ GeV, while in the $Z \to (\gamma \gamma)\gamma$ decay we further require the third photon form the $Z$ resonance together with the first two photons~\cite{Liu:2017zdh}. We summarize the basic cuts and the specific requirements for each exotic $Z$ decay channel in the following,
\begin{eqnarray}
	& \bullet & \quad \text{Basic cuts}: E_\gamma>2~\text{GeV}, p_T^\mu > 0.1~\text{GeV}, ~|\eta_{\gamma/\mu}|<3.0 \nonumber \\ 
	& \bullet &\quad  Z \to (\mu^+ \mu^-) \gamma: \left| \frac{m_{\mu\mu} -m_a}{m_a} \right| < 0.19\% , \nonumber\\
	& \bullet & \quad  Z \to (\gamma_1 \gamma_2) \gamma_3: \left|m_{\gamma_1 \gamma_2} -m_a \right| < 1~{\rm GeV}, 
	~ \left|E_{\gamma_3} - \frac{m_Z^2 -m_a^2}{2 m_Z} \right| < 1~{\rm GeV}, \nonumber\\
	& \bullet & \quad  Z \to (\mu^+ \mu^-) \mu^+ \mu^- : \text{at least one opposite sign muon pair} \left| \frac{m_{\mu\mu} -m_a}{m_a} \right| < 0.19\% ,\nonumber\\
	&\quad &  \quad \Delta R_{\mu^+\mu^-} > 0.1,  ~p_T^\mu > 5~\text{GeV}.
	\label{eq:cuts}
\end{eqnarray}
For Higgs factory with $\sqrt{s}=240$ GeV, we substitute $m_Z$ in the above cuts to $240$ GeV.
The main results of our study, with the assumption of $C_{WW}=0$, are shown in Fig.~\ref{fig:ma-varying-final-plot} and  Fig.~\ref{fig:finalresults}.

\subsection{Summary of current limits and projected reach, $C_{WW} = 0$.}

Based on the discussions in Section~\ref{sec:existConstraints} and Section~\ref{sec:futureZfactory}, we summarize the current constraints and the projected reaches of Z factories (CEPC and FCC-ee). 

\begin{figure}[htbp]
	\includegraphics[width=0.32\linewidth]{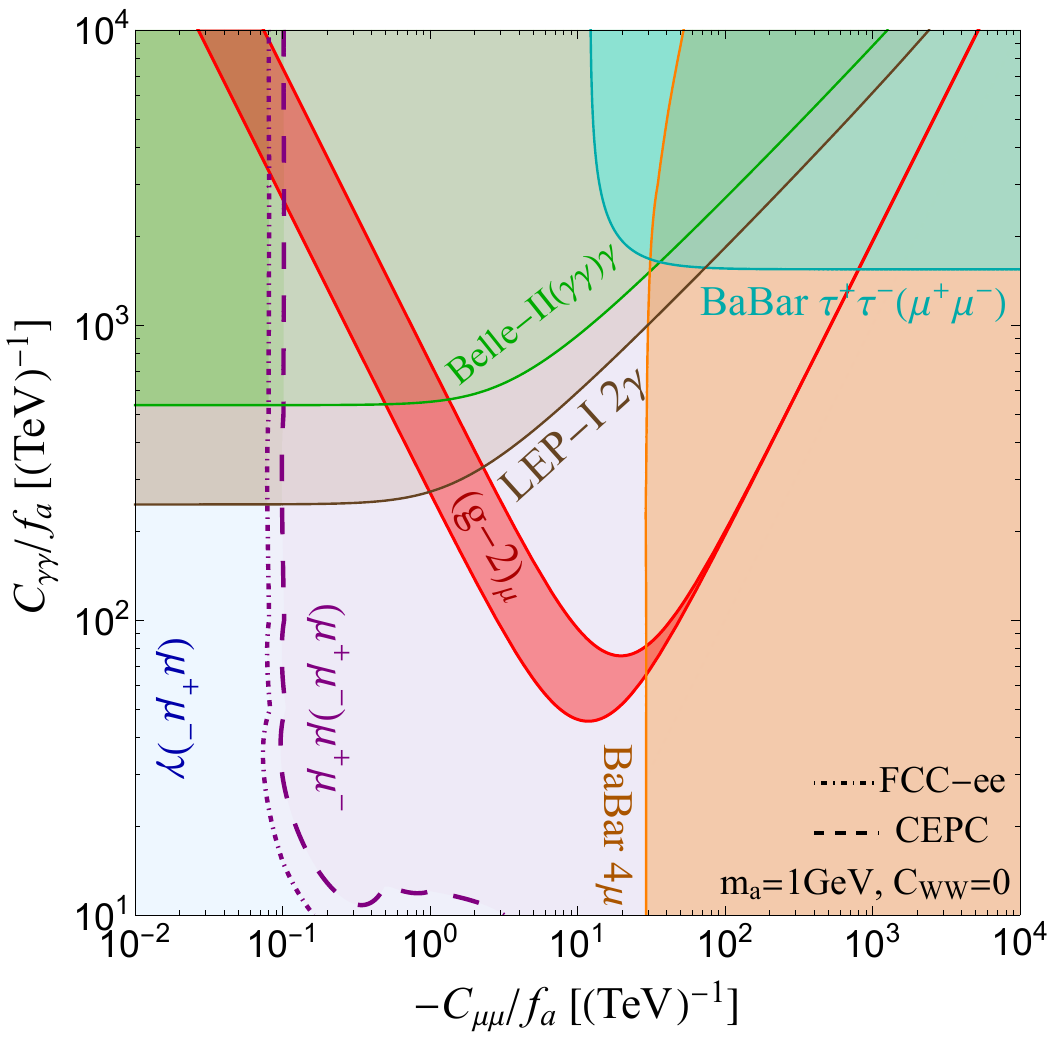}
	\includegraphics[width=0.32\linewidth]{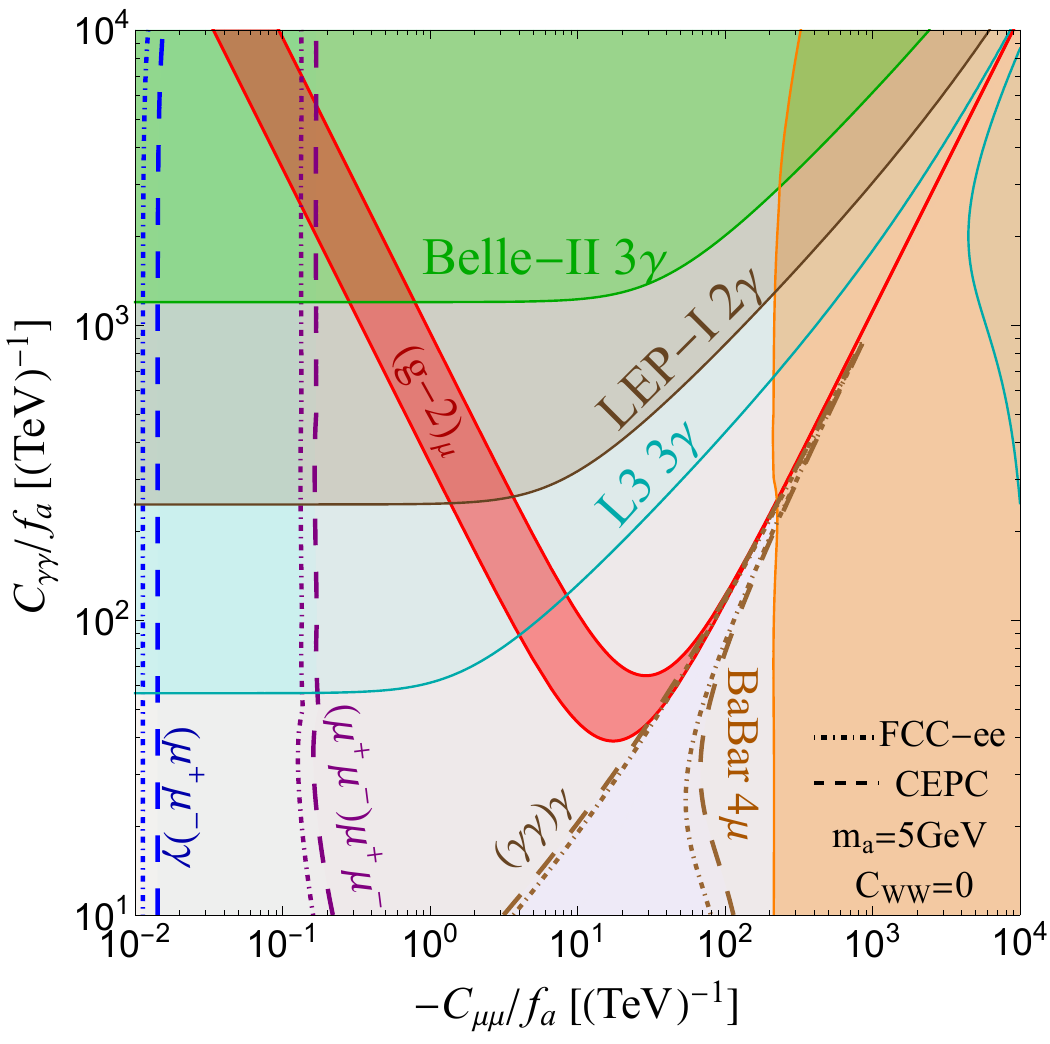} 
	\includegraphics[width=0.32\linewidth]{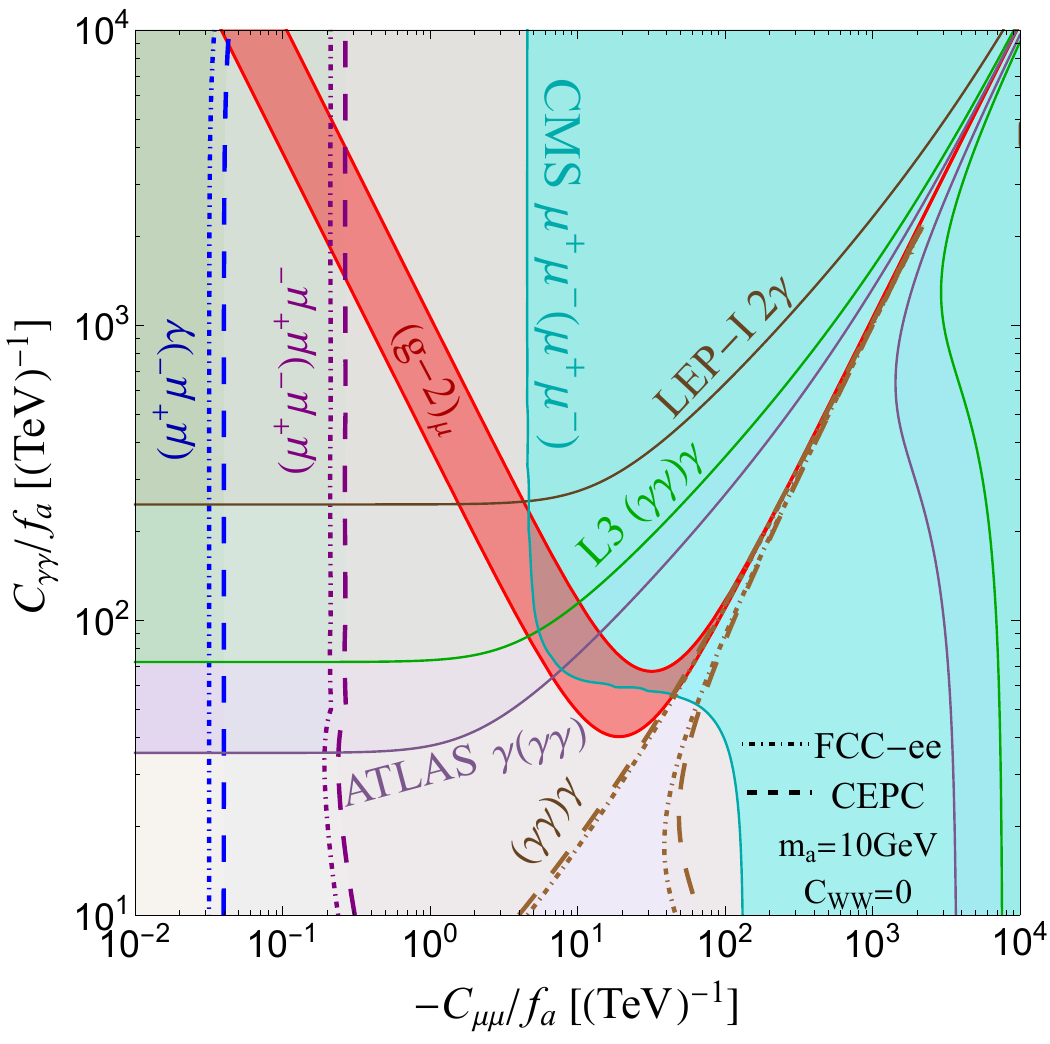}\\
	\includegraphics[width=0.32\linewidth]{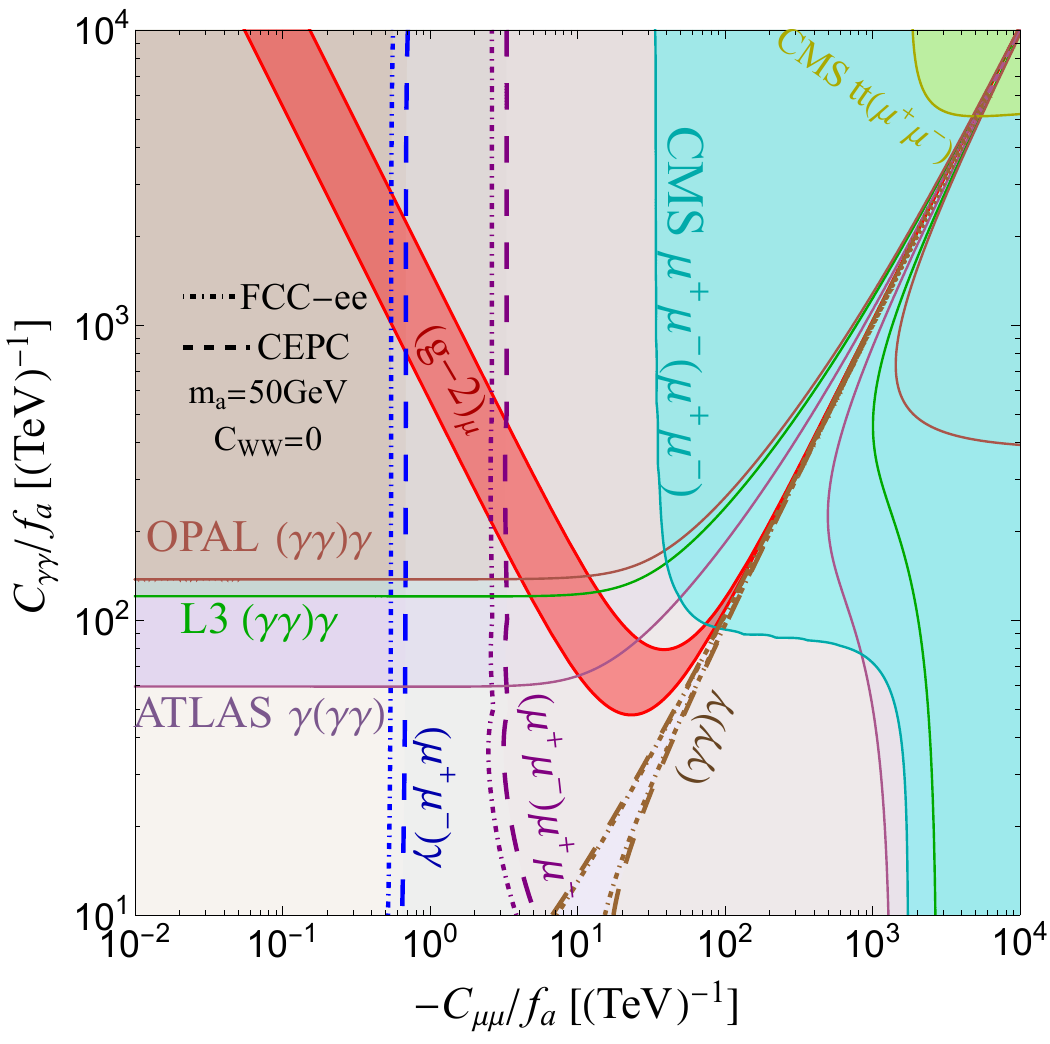}
	\includegraphics[width=0.32\linewidth]{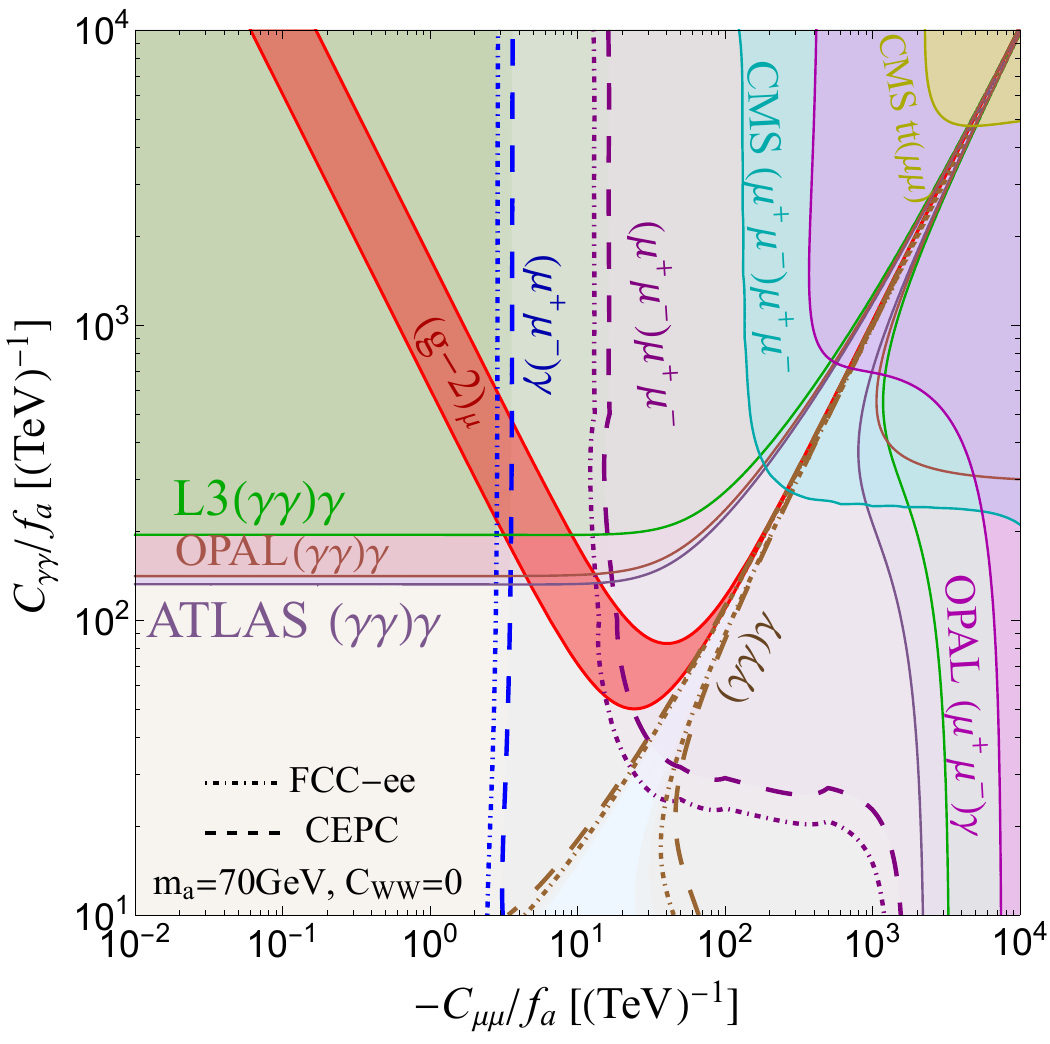} 
	\includegraphics[width=0.32\linewidth]{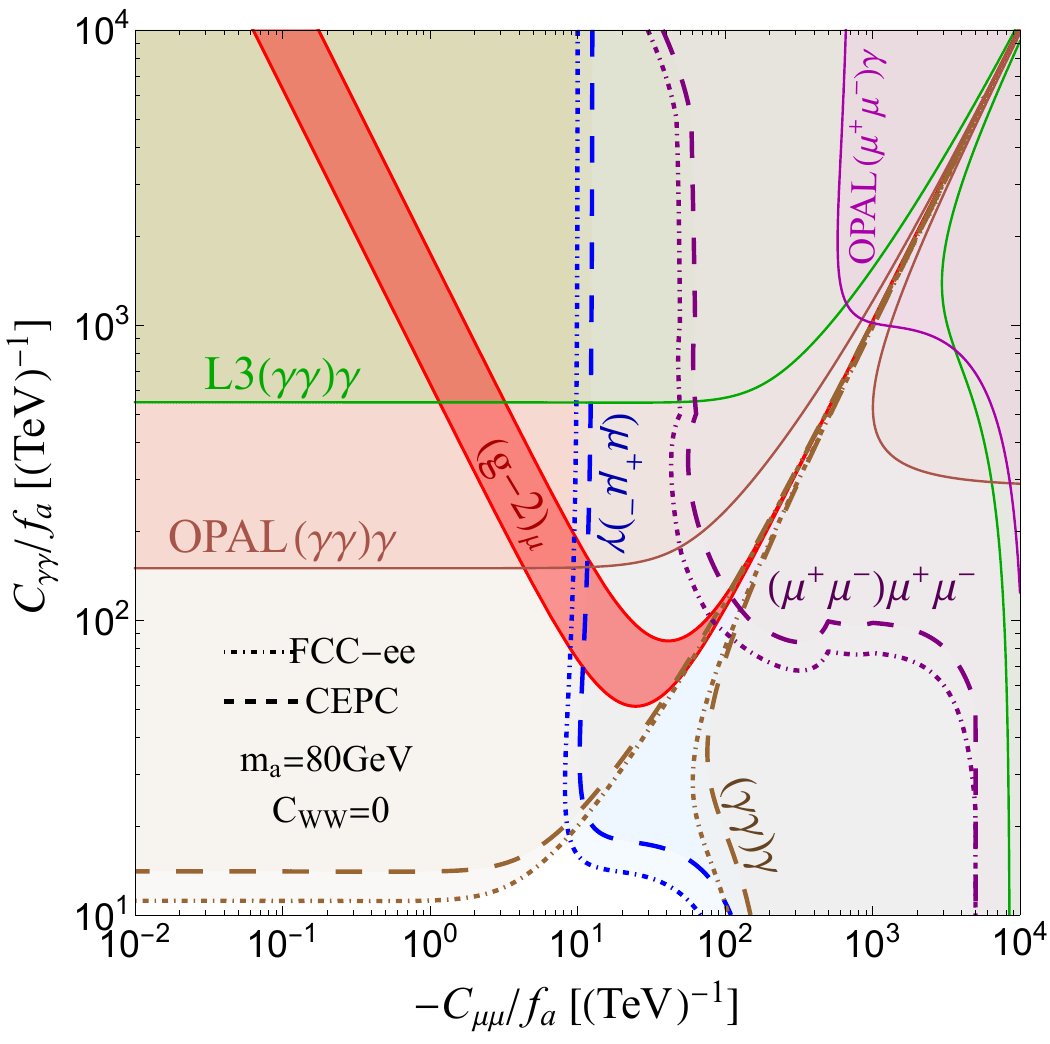}
	\caption{The constraints from various existing experiments and future projections from $Z$ factories (CEPC and FCC-ee) from exotic $Z$ decay searches
		$Z \to a \gamma \to (\gamma\gamma)\gamma$, $Z \to a \gamma \to (\mu^+ \mu^-)\gamma$ and $Z \to a \mu^+ \mu^- \to (\mu^+ \mu^-)\mu^+ \mu^-$.
		The parameter space of $(g-2)_\mu$ solution is plotted in the red band.
		We set $C_{WW}=0$, and $\Lambda=1~\text{TeV}$ in $g-2$ calculation. }
	\label{fig:ma-varying-final-plot}
\end{figure}

Fig.~\ref{fig:ma-varying-final-plot} shows the result on the ALP couplings $C_{\mu \mu}/f_a$ and $C_{\gamma \gamma}/f_a$,  for different benchmark ALP masses. In particular, the results for $m_a = 1,~5,~10,~50,~70,$ ~and 80 GeV are plotted. 
For light $m_a$ and small  $C_{\mu\mu}$, the  parameter region capable of explaining the $(g-2)_\mu$ anomaly (red band) is constrained by $a\gamma \to (\gamma\gamma)\gamma$ from Belle-II~\cite{Belle-II:2020jti}, L3~\cite{L3:1994shn}, OPAL~\cite{OPAL:2002vhf} and ATLAS~\cite{ATLAS:2015rsn} or inclusive $\gamma\gamma$ from LEP-I~\cite{Jaeckel:2015jla} searches. At the same time, for large $C_{\mu\mu}$, the decay channel to $a\to \mu^+\mu^-$ leads to constraints from BaBar $(\mu^+\mu^-)\mu^+\mu^-$
and $\tau^+\tau^- a \to \tau^+\tau^- (\mu^+\mu^-)$ searches~\cite{BaBar:2020jma,BaBar:2016sci}. 
For large $m_a$ mass, the similar feature holds that the small $C_{\mu\mu}$ is constrained by $3\gamma$ or $2\gamma$ searches, while large $C_{\mu\mu}$ is constrained by axion muonic decay searches.
In contrast to the low $m_a$ case, the CMS searches of $pp\to Z \to \mu^+\mu^- \phi$ and $pp\to \bar{t}t \phi$~\cite{CMS:2019lwf,CMS:2018yxg} with $\phi$ decaying to muon pair come into play due to invariant mass threshold. The former requires the reconstruction of the $Z$ boson from four muon invariant mass. The latter does not require $\phi$ mass to be around $m_Z$,  and its limit for $\phi$ mass higher than $Z$ is too weak to show up on the figure. The OPAL search $Z \to \mu^+ \mu^- \gamma$~\cite{OPAL:1991acn} is also relevant and complements the CMS limits. 

For the range of $m_a$ considered here, the limits from $3\gamma/2\gamma$ final states disappear for the right part of the $(g-2)_\mu$ band, where $C_{\gamma \gamma} + C_{\mu\mu} \sim 0$. This is because the ${\rm BR}(a\to \gamma \gamma)$ is governed by $C_{\gamma\gamma}^{\rm eff}$ in Eq.~\ref{eq:C_AA_eff}, which is vanishingly small in this case. 
This region is easily covered by $Z$-factory searching for $Z \to (\mu^+\mu^-)\gamma$ and $Z \to (\mu^+\mu^-)\mu^+\mu^-$, which benefit from the vanishing ${\rm BR}(a\to \gamma \gamma)$.
Together with $Z \to (\gamma\gamma)\gamma$, the $Z$-factory can cover the rest of the parameter space relevant for an explanation of the $(g-2)_\mu$ anomaly up to $m_a \lesssim 85$ GeV, providing a decisive check for the ALP solution to $(g-2)_\mu$ and is complementary to other existing experiments.

\begin{figure}[htbp]
	\centering
	\includegraphics[width=0.48 \textwidth]{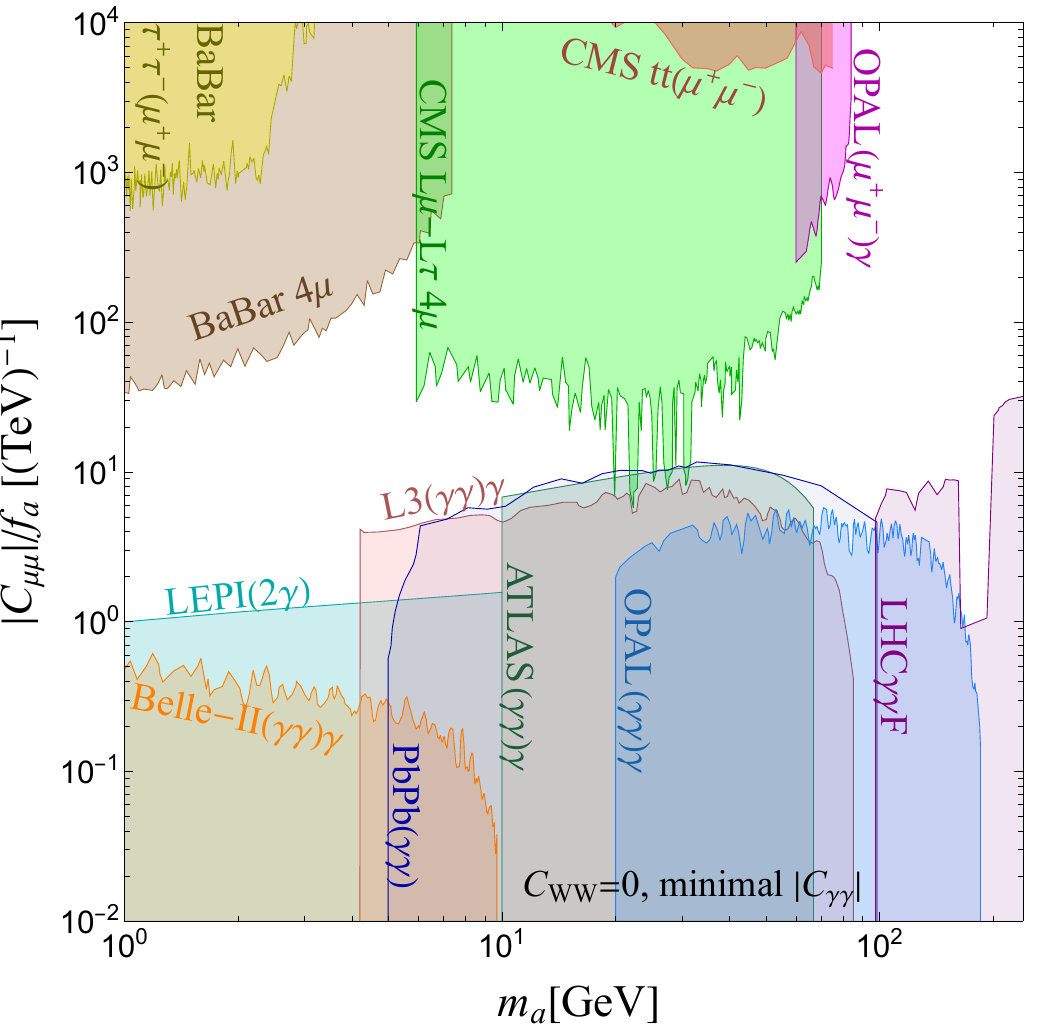}
	\includegraphics[width=0.48 \textwidth]{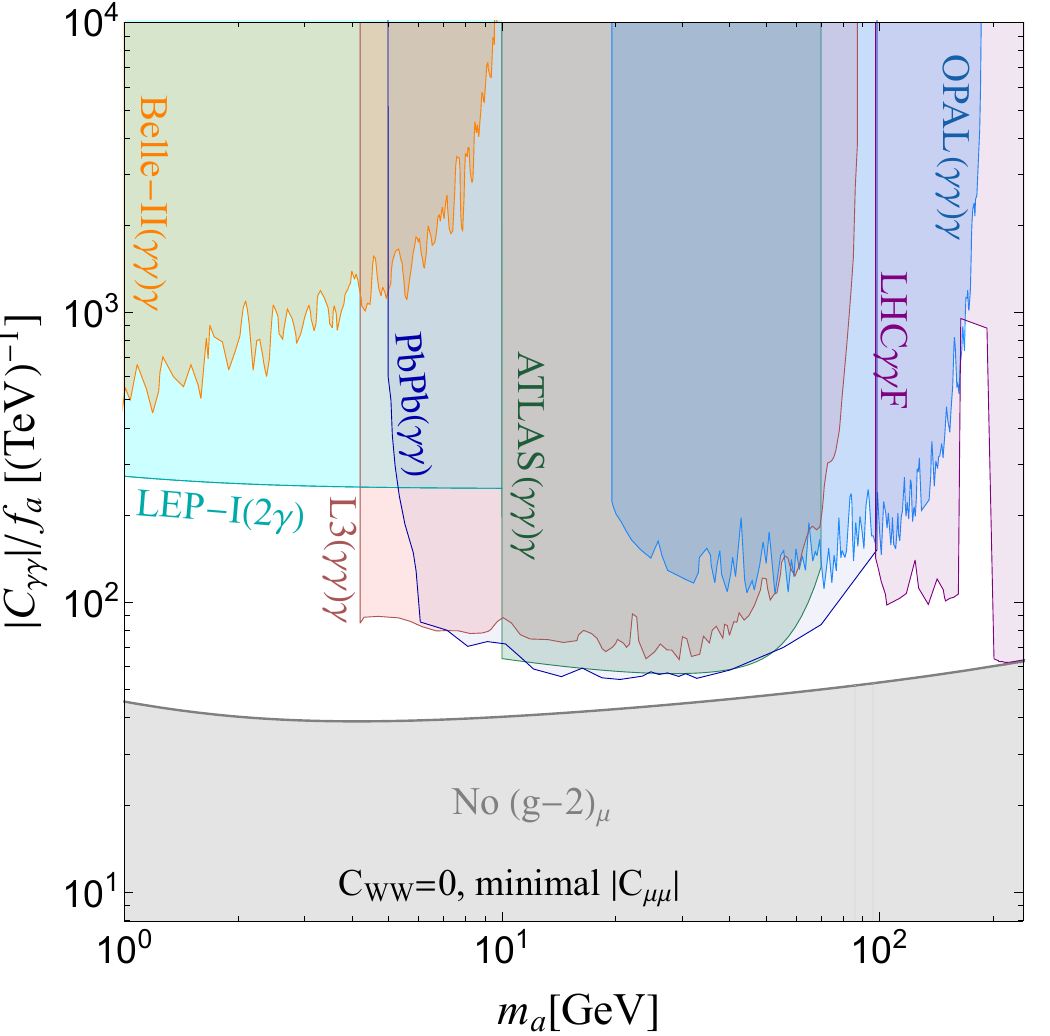} \\
	\includegraphics[width=0.48 \textwidth]{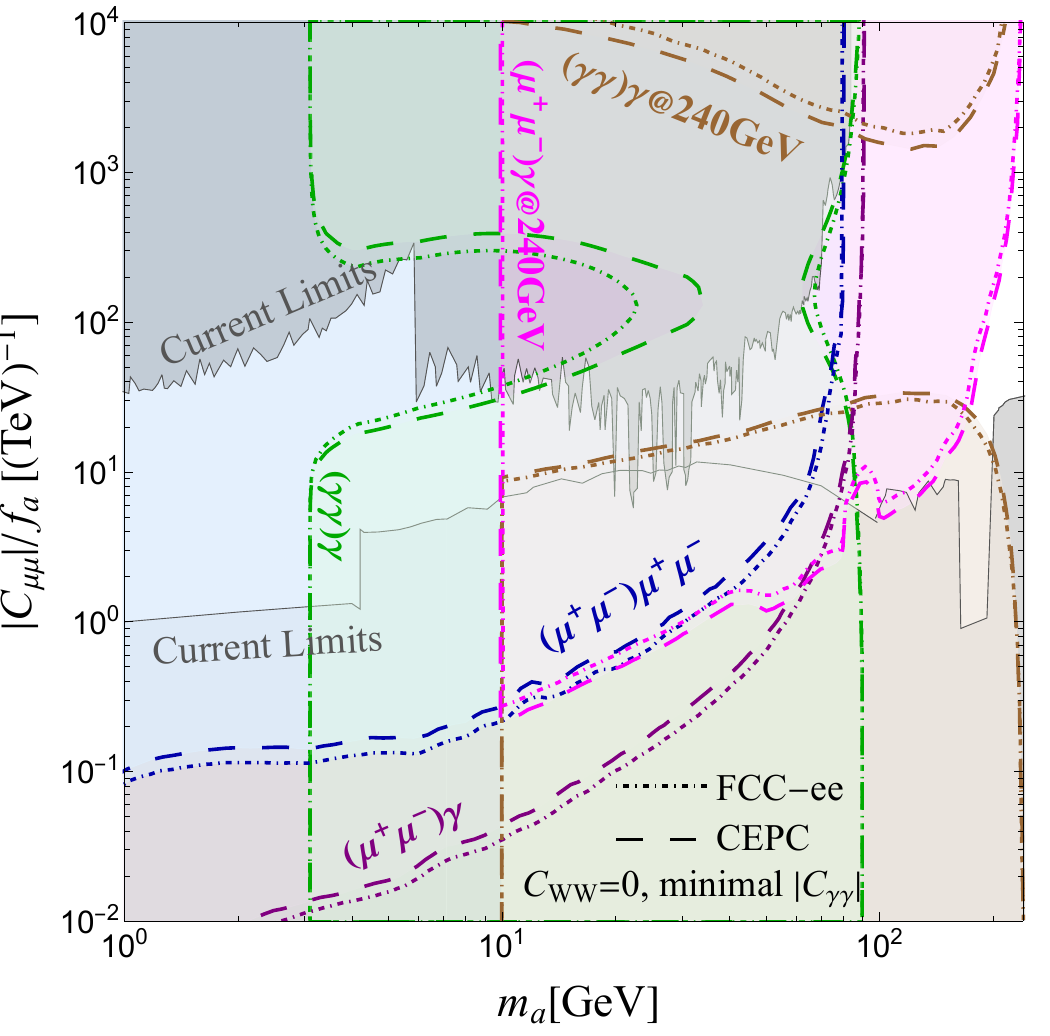}
	\includegraphics[width=0.48 \textwidth]{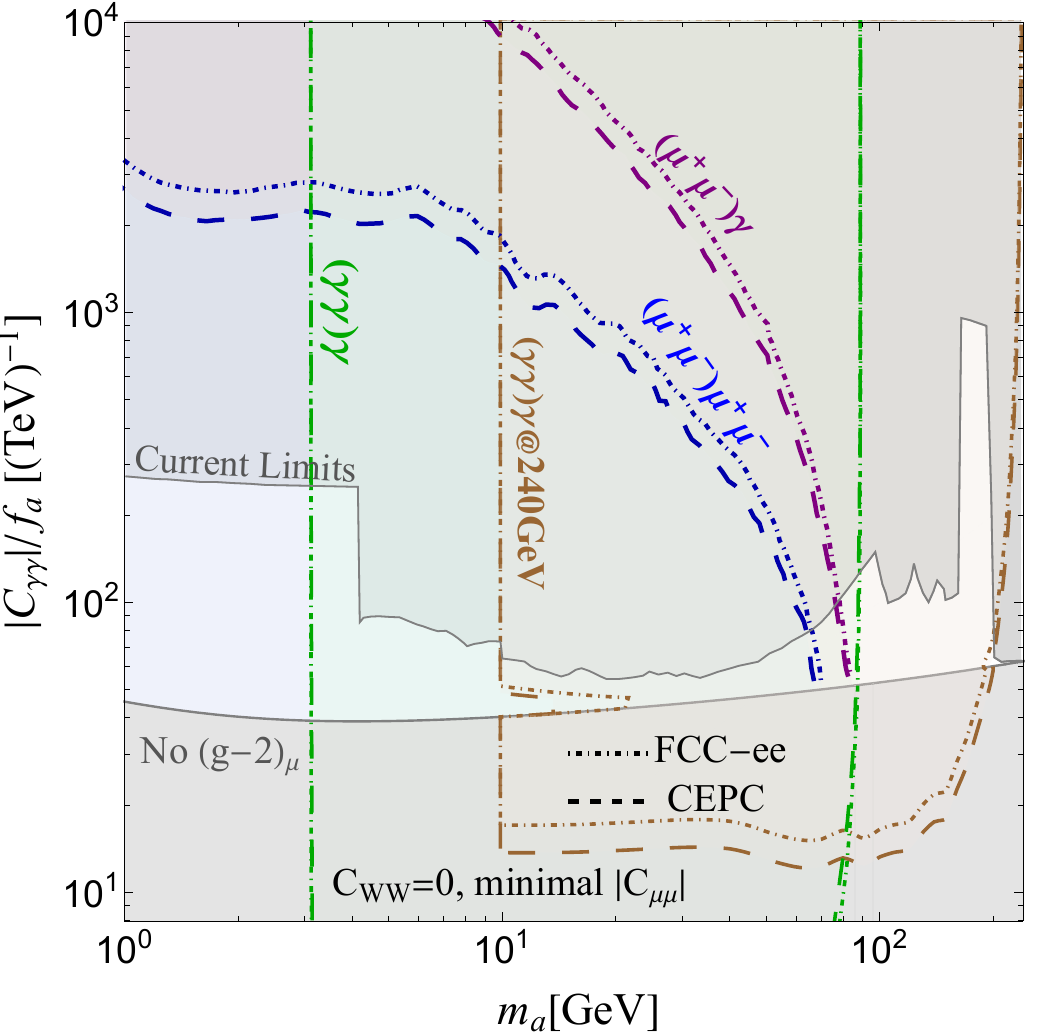}
	\caption{
		\textit{Left panels}: The existing constraints and future sensitivities in the  $C_{\mu\mu}/f_a-m_a$ plane. We set $C_{WW}=0$.  
	For each pair of parameters,  $C_{\mu\mu}$ and $m_a$, we choose the minimal $|C_{\gamma\gamma}|$ which can explain the $(g-2)_\mu$ anomaly at $2\sigma$ level.
	\textit{Right panels}: The existing constraints and future sensitivities in the  $C_{\gamma\gamma}/f_a-m_a$ plane with $C_{WW}=0$.
	We choose the minimal $|C_{\mu\mu}|$ coupling which can satisfy $(g-2)_\mu$ anomaly at $2\sigma$ level.
	The gray region has no viable $g-2$ solution and we simply set $C_{\mu\mu}=0$ here.
	\textit{Top panels}:
	The existing constraints are plotted in color with the solid line as a boundary at $95\%$ confidence level. 
	\textit{Bottom panels}:
		The reaches of future $Z$-factories at CEPC (dashed lines) and FCC-ee (dot-dashed lines) with searches $(\gamma \gamma)\gamma $, $(\mu^+ \mu^- )\gamma $ and $(\mu^+ \mu^- ) \mu^+ \mu^-$ are shown, while the reaches of the future Higgs factories are labeled with an extra $@ 240 {\rm GeV}$.}
	\label{fig:finalresults}
\end{figure}

Fig.~\ref{fig:finalresults} presents the existing constraints and future reaches in the $C_{\mu\mu}(C_{\gamma\gamma})/f_{a} - m_a$ plane, where we place the existing ones in the top panel and future ones in the bottom panel.
To focus further on the relevant part of the parameter region, we will impose the condition that the parameter which is not plotted is chosen so that an explanation of the $(g-2)_\mu$ anomaly is possible. 
In the left panel, we choose $C_{\mu\mu}$ and $m_a$ as free parameters,
while  $C_{\gamma \gamma}$ is chosen to be the minimal value which can give an explanation to the $(g-2)_\mu$ anomaly at $2\sigma$ level. In the right panel,  we choose the minimal $C_{\mu\mu}$ in a similar way. The only exception is the gray region in the right panel where a possible explanation for the $(g-2)_\mu$ anomaly can not be found within the model under consideration.

In the left panel of Fig~\ref{fig:finalresults}, one might expect that a larger $C_{\gamma\gamma}$ could help to evade the $a\to \mu^+ \mu^-$ searches via smaller branching ratio. However, this effect is compensated by the increased cross-section of  $e^+e^- \to a \mu^+ \mu^-$
and $Z \to a \mu^+ \mu^-$, because large $C_{\gamma\gamma, \gamma Z}$ will dominate the contribution comparing with $C_{\mu\mu}$. 
This feature is clear in Fig.~\ref{fig:ma-varying-final-plot} that the BaBar $4\mu$ and $Z$-factory $a(\mu^+ \mu^-)\gamma, ~a(\mu^+ \mu^-)\mu^+ \mu^-$ boundaries become vertical with increasing $C_{\gamma\gamma}$.
The existing constraints leave the lower half of the plot unconstrained. 
At the same time, the future probes from $Z$-factories can fully cover the parameter space up to $m_a \sim 85$ GeV.

The constraints from photon final states $(\gamma\gamma) \gamma$ set limits on the small $C_{\mu \mu}$ region, where a large $C_{\gamma \gamma}$ is necessary for explaining the $(g-2)_\mu$ anomaly. The constraints from low energy $e^+ e^-$ collider Belle-II~\cite{Belle-II:2020jti},   L3~\cite{L3:1994shn} search based on $Z$-pole data, OPAL~\cite{OPAL:2002vhf} search based on the data collected during the $\sim 200$ GeV run, pp collider $Z$ study from ATLAS~\cite{ATLAS:2015rsn}, photon fusion from PbPb ion collision~\cite{CMS:2018erd,ATLAS:2020hii} and pp collision at LHC~\cite{CMS:2021plq,CMS:2012cve} are shown in the top-left panel.
Together with $\mu^+\mu^-$, it still leaves the possibilities open for  $|C_{\mu\mu}|/f_a \sim 10 ~{\rm TeV}^{-1}$ for $m_a < m_Z$ and for $m_a > m_Z$ with large $C_{\mu \mu}$. Adding the constraints from exotic $Z$ decay searches $Z$-factory, one can fully cover the parameter space for $m_a < 85$ GeV. With the $240$ GeV run, one can extend the exclusion up to $m_a \sim 160$ GeV via $e^+e^- \to (\gamma \gamma)\gamma$ and $(\mu^+ \mu^-)\gamma$, which is complementary to the LHC studies. It can cover part of parameter region for $m_a$ up to 240 GeV, but leaves a small opening for $|C_{\mu\mu}|/f_a \sim 10$--$100~{\rm TeV}^{-1}$. 

 In the right panel of Fig.~\ref{fig:finalresults}, for $C_{\gamma\gamma}$ within the ``No $(g-2)_\mu$" band (gray region), there is no viable solutions and we set $C_{\mu \mu }= 0$ in this region by hand. 
The existing searches for $3\gamma/2\gamma$ multi-photon states have excluded a good portion of the parameter space in the top-right panel. The OPAL $(\gamma \gamma) \gamma$, photon fusion at PbPb ion collision and pp collider searches at high energy can cover ALP masses larger than $m_Z$ because they do not rely on the on-shell $Z$.
They  leave an open space for $|C_{\gamma\gamma}|/f_a \sim 50$--$200 ~{\rm TeV}^{-1}$.
The future $Z$-factory probes can cover the parameter region for the  ALP explanation to $(g-2)_\mu$ anomaly up to $m_a = $85 GeV. 

The $(\gamma \gamma) \gamma$ search from future Higgs factories can cover the higher mass region up to $m_a \sim 160$ GeV. One can see that the relevant parameter space is almost excluded. 

\subsection{Summary of constraints and projected reaches, $C_{WW} \neq 0$.}
\label{sec:generalGaugeCouplings}
Our results on the ALP contribution to $(g-2)_\mu$ can be extended to cases with $C_{WW}\neq 0$ in a straightforward way. In this case, the couplings $ C_{\gamma \gamma}$ and $C_{\gamma Z} $ are independent. 
There are four free parameters for the model,
$ 	 m_a,~C_{\mu\mu},~C_{WW}$ and $C_{BB} $.
Since the 2-loop Barr-Zee diagram with $W$ boson running in the loop is small compared with other diagrams, the $C_{WW}$ contribution to $(g-2)_\mu$ comes mainly from the fact that it enters the independent $ C_{\gamma \gamma}$ and $C_{\gamma Z} $ couplings. 

\begin{figure}[htbp]
	\includegraphics[width=0.32\linewidth]{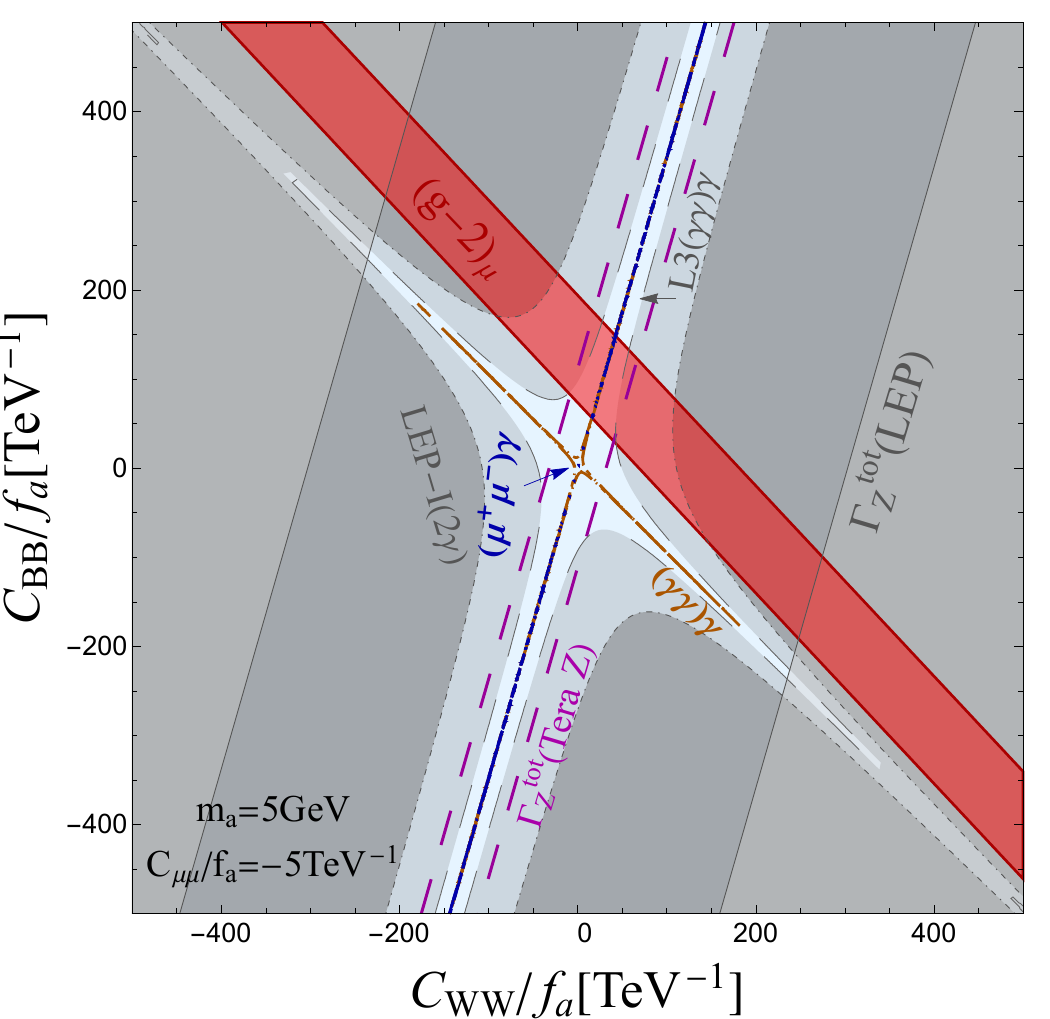}
	\includegraphics[width=0.32\linewidth]{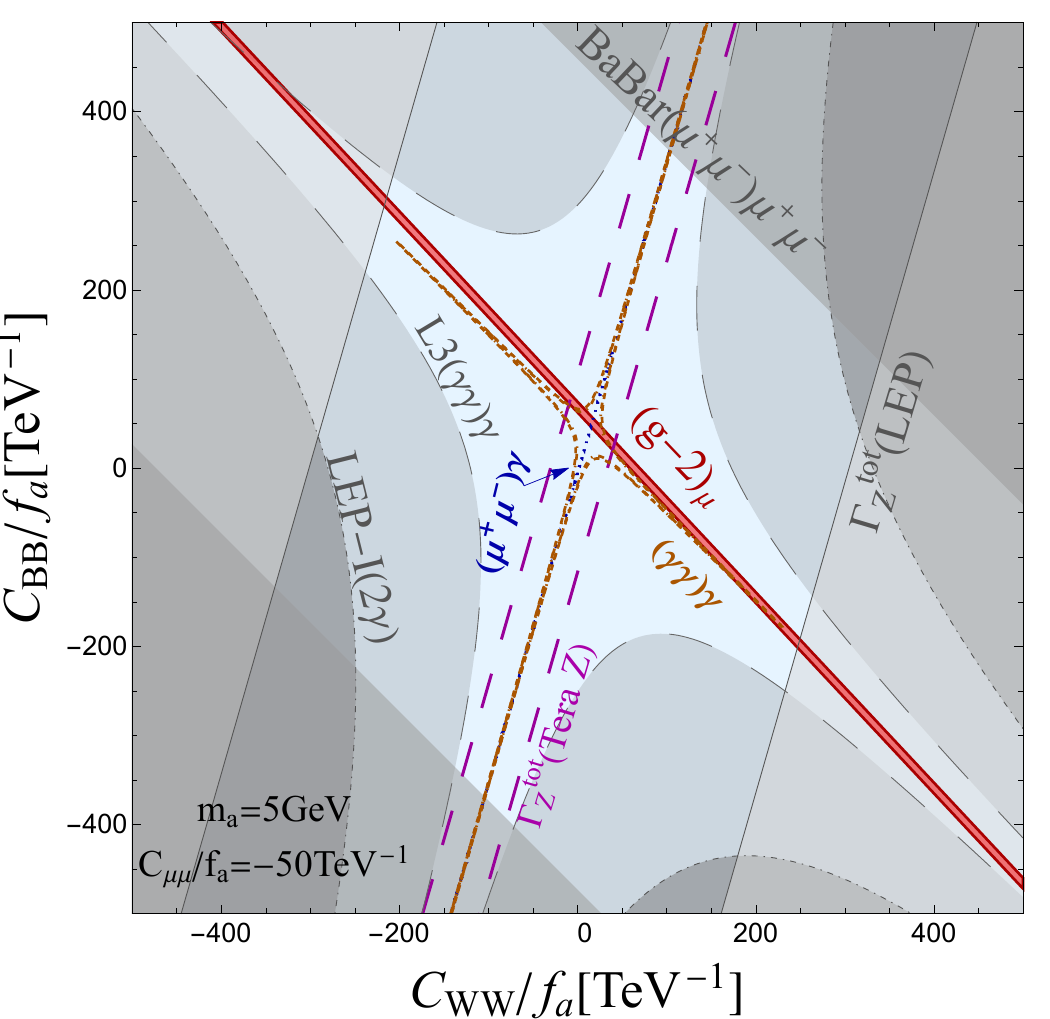}
	\includegraphics[width=0.32\linewidth]{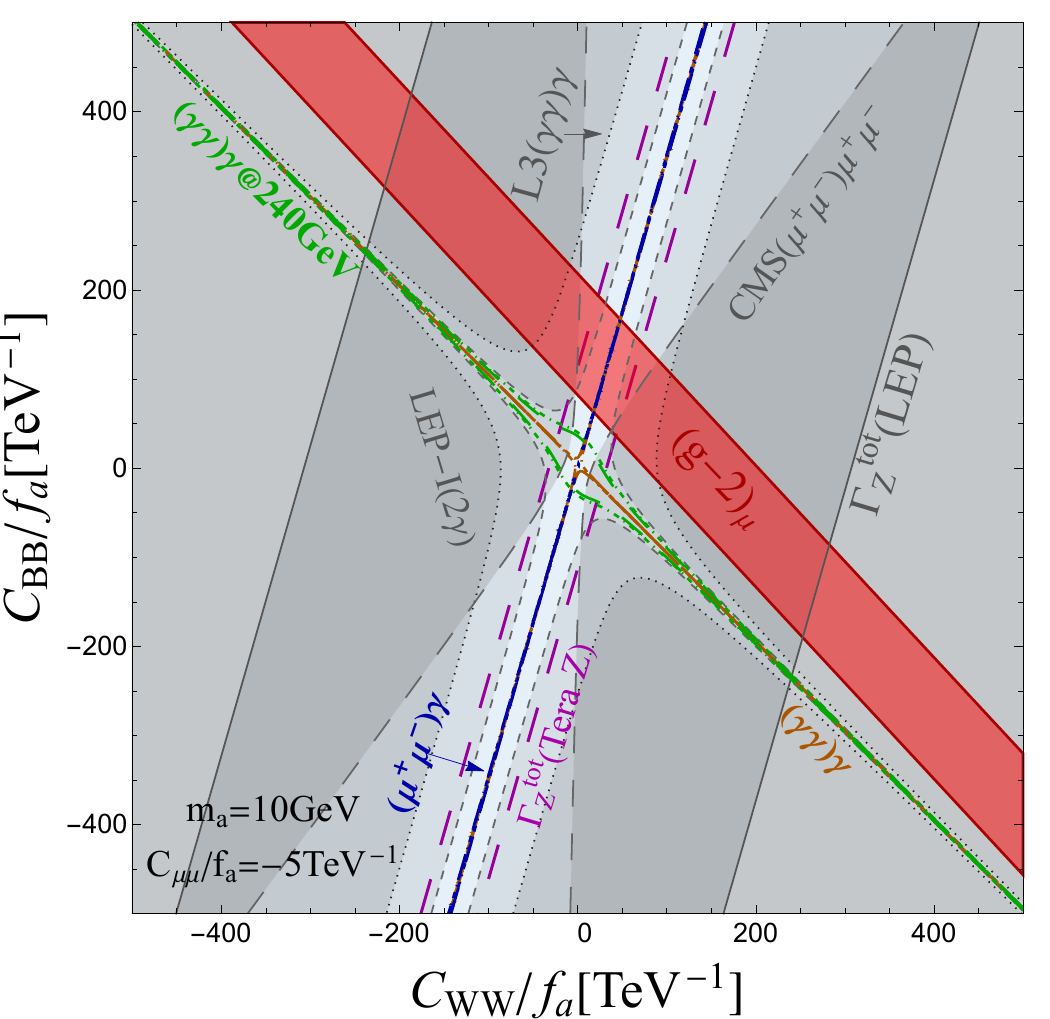} \\
	\includegraphics[width=0.32\linewidth]{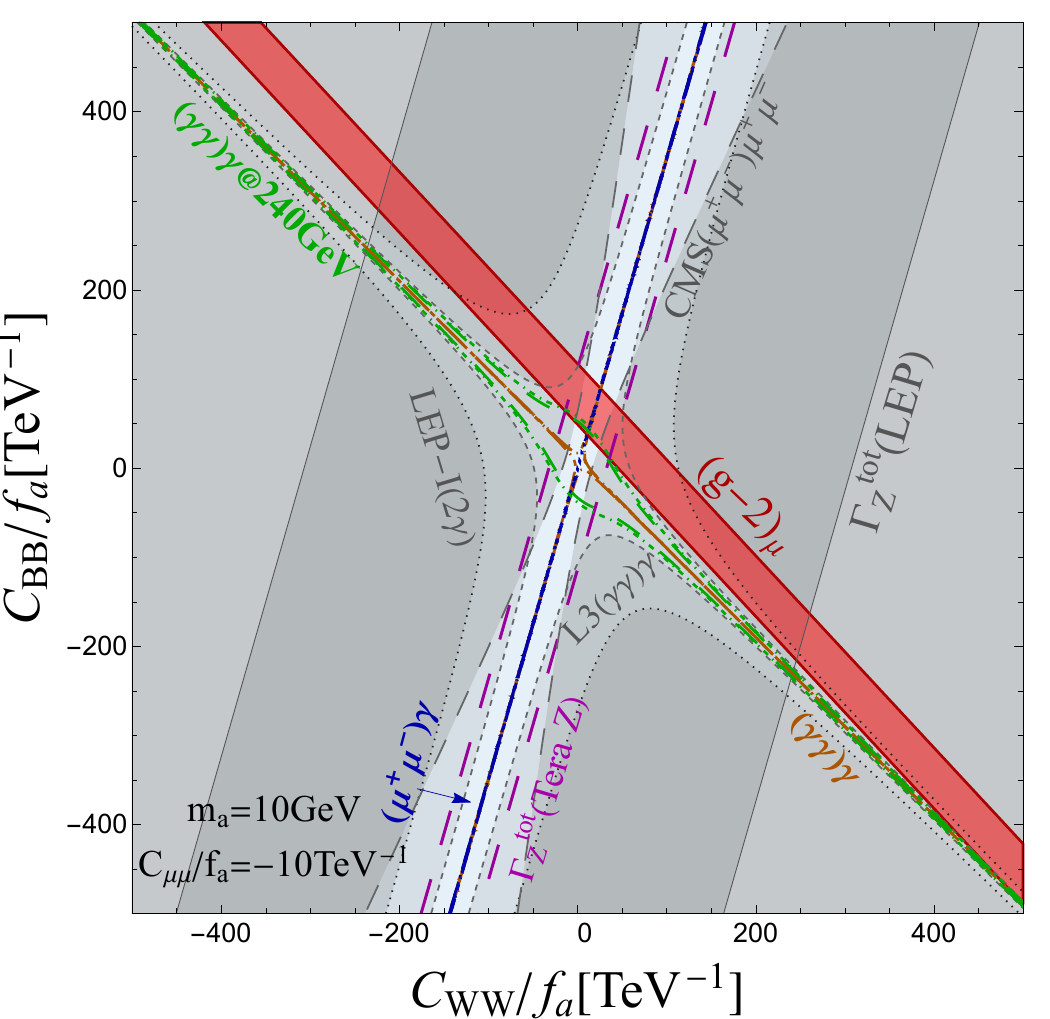} 
	\includegraphics[width=0.32\linewidth]{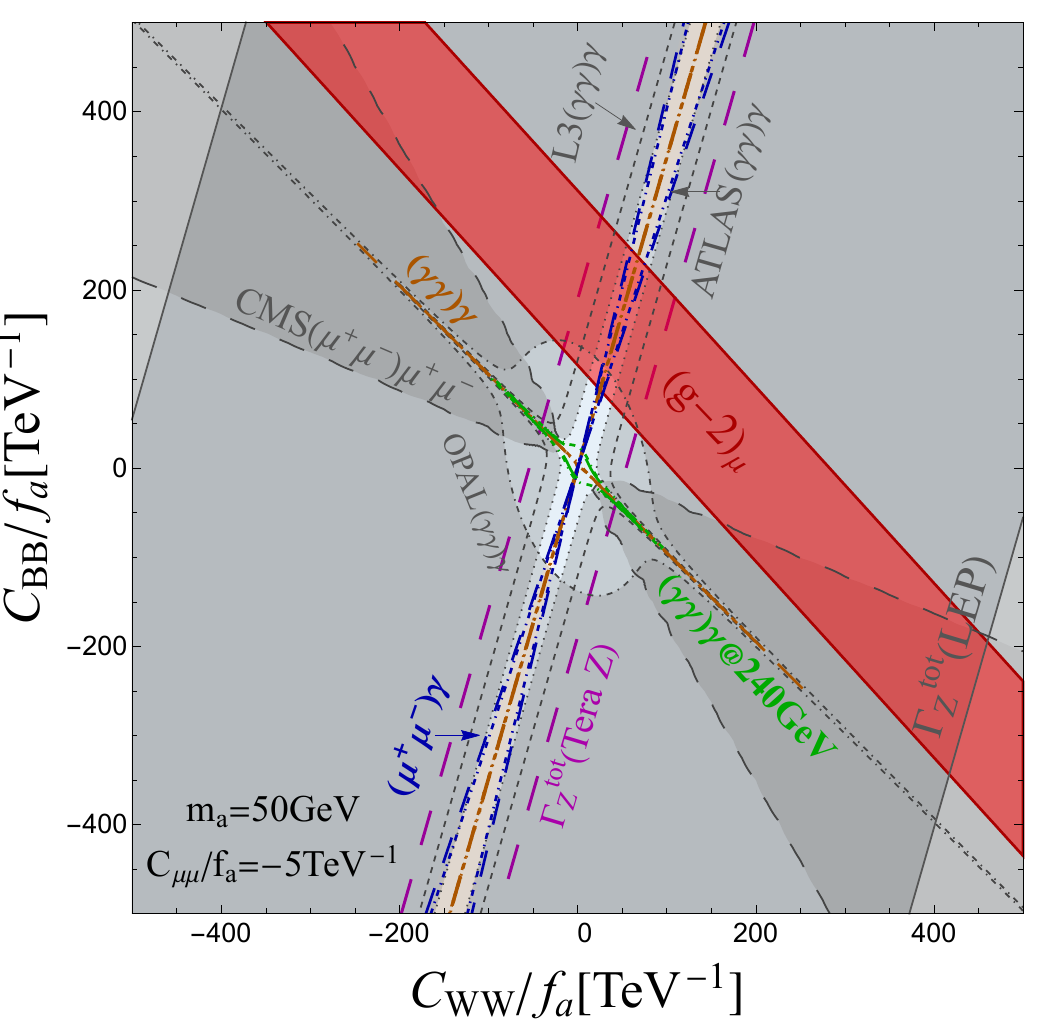} 
	\includegraphics[width=0.32\linewidth]{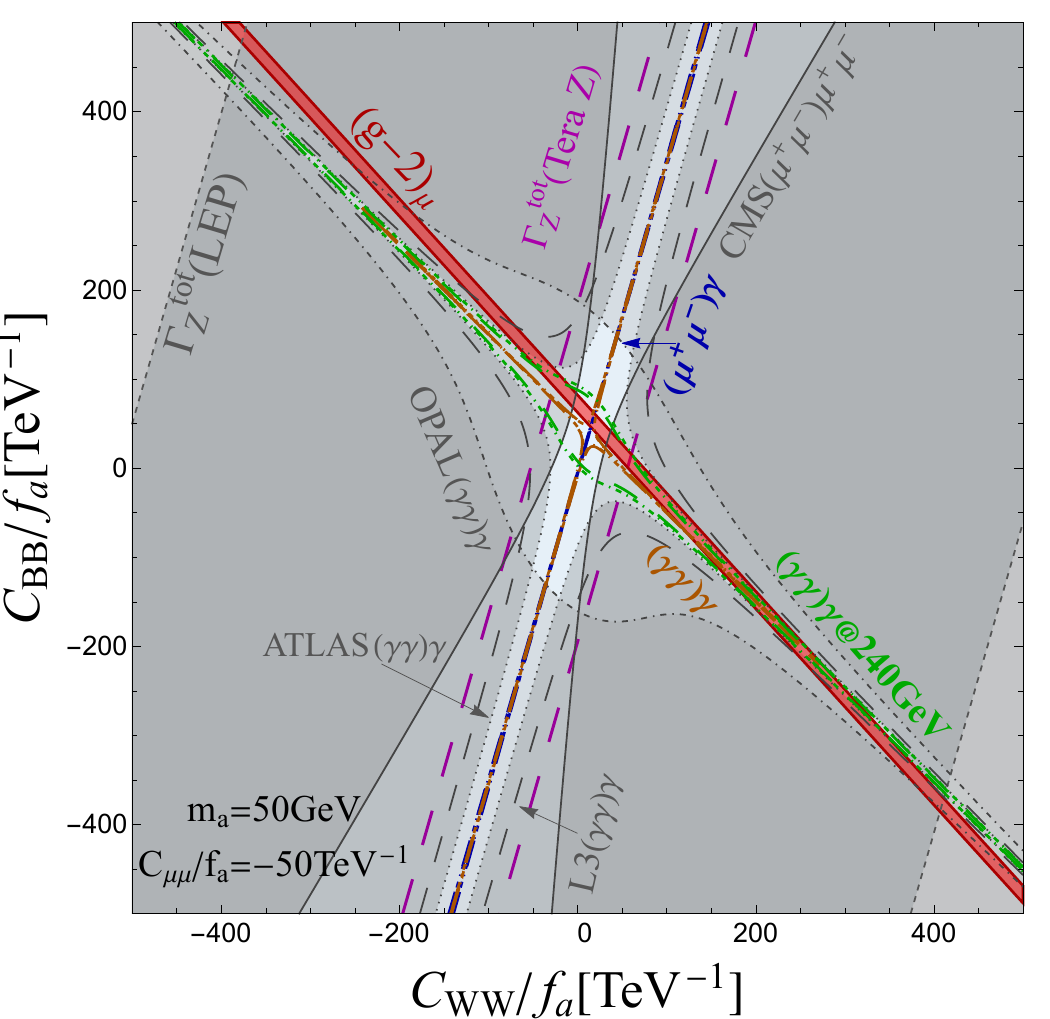}\\
	\includegraphics[width=0.32\linewidth]{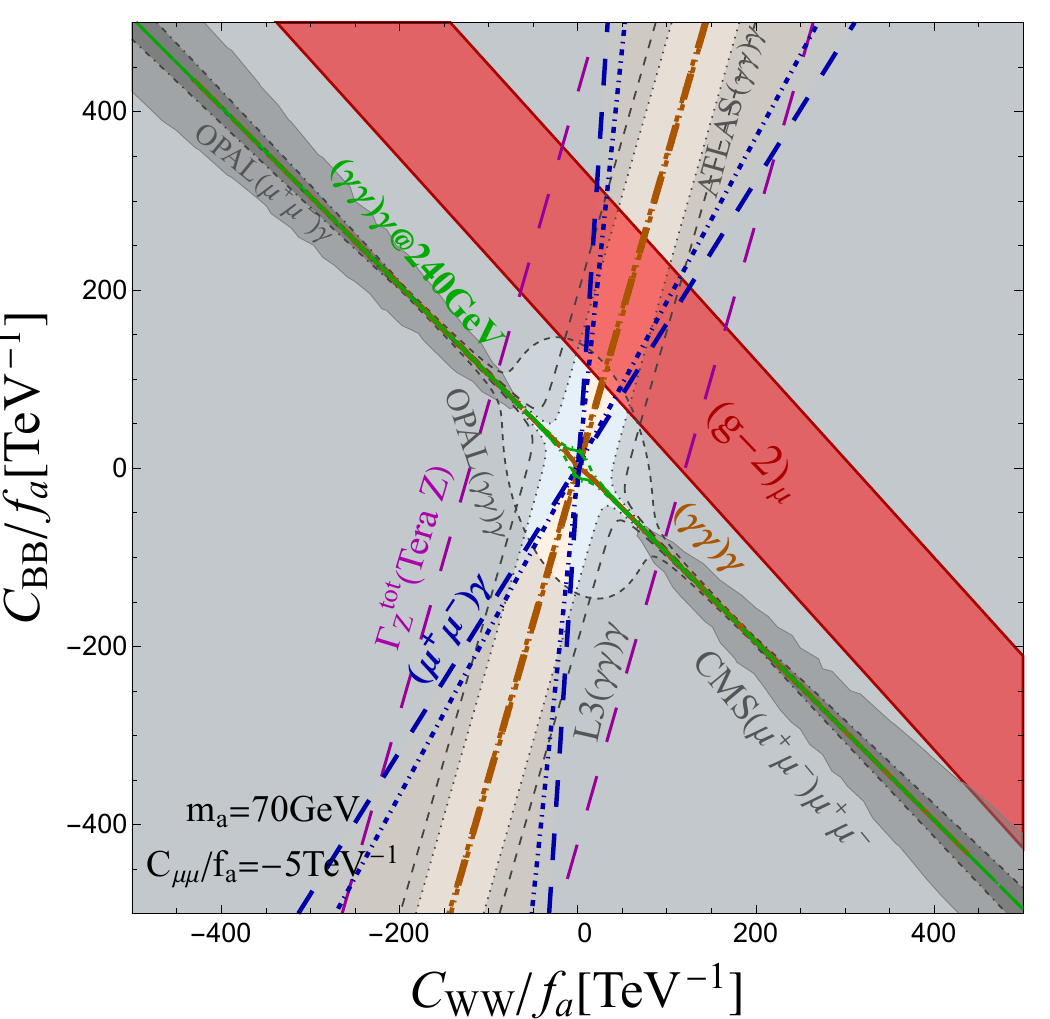}
	\includegraphics[width=0.32\linewidth]{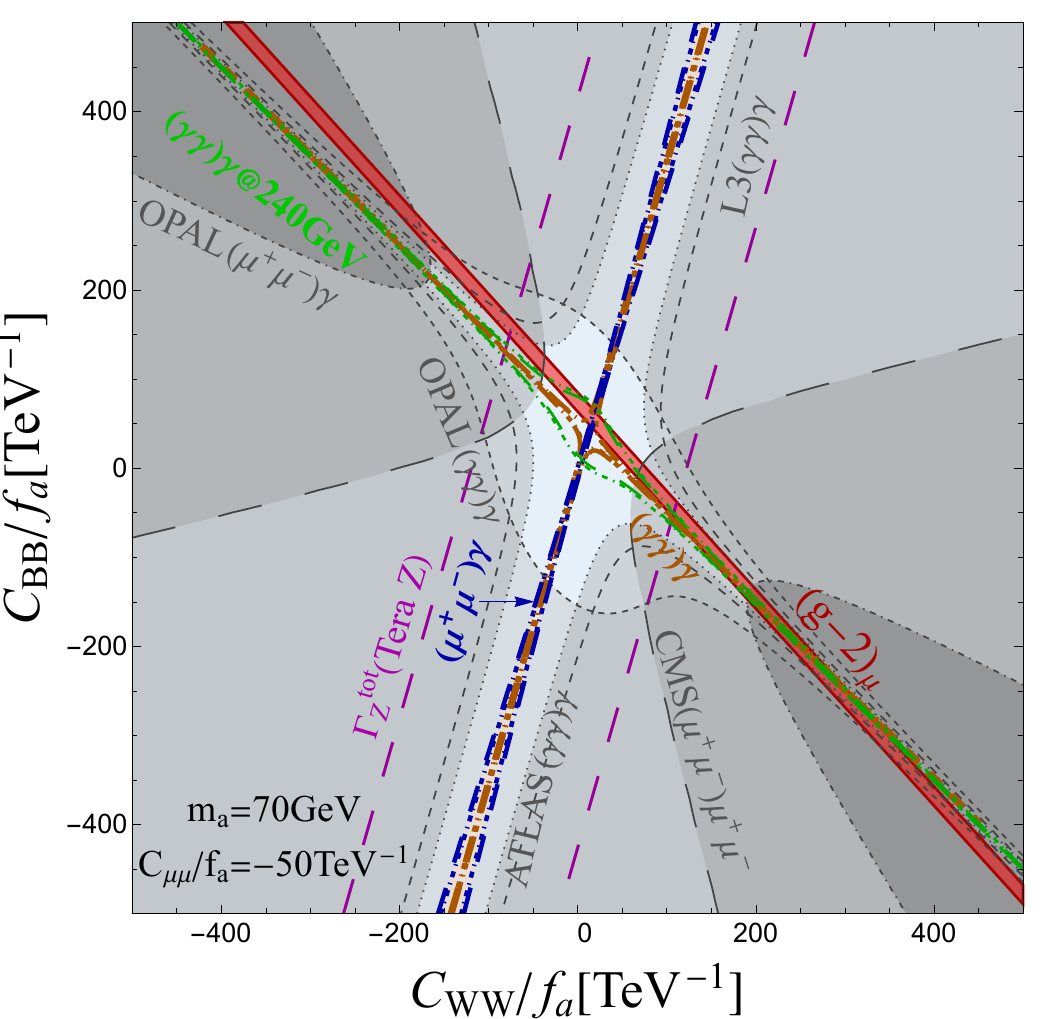}
	\includegraphics[width=0.32\linewidth]{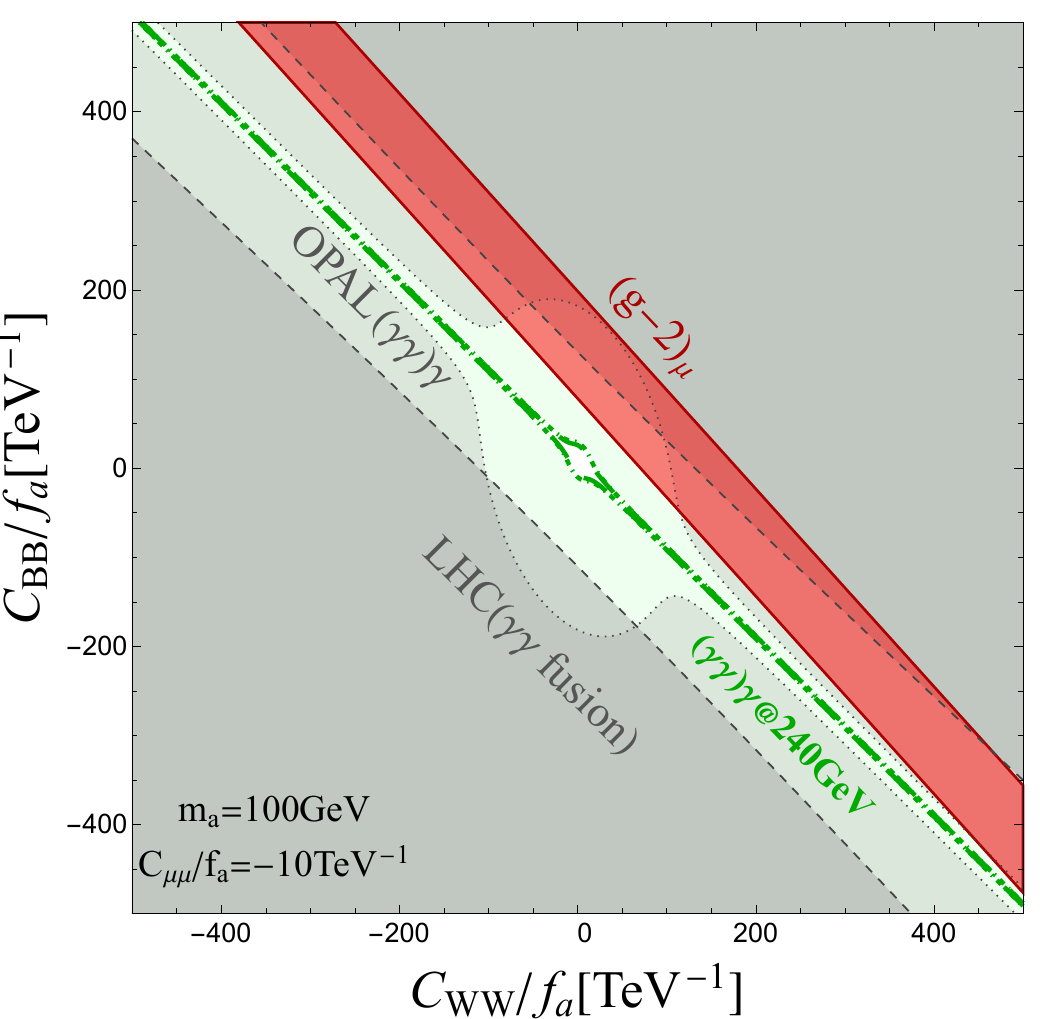}
	\caption{The existing constraints and future sensitivities from $Z$-factory with $C_{WW} \neq 0$. The color codes are similar to Fig.~\ref{fig:finalresults}. We choose $m_a =5,~10,~50,~70,~100$ GeV respectively together with  appropriate small muon couplings $C_{\mu\mu}$.}
	\label{fig:general1}
\end{figure}

Since we have more free parameters,  it becomes difficult to present the full results in two-dimensional plane. Instead,  it is illuminating to present the results in $C_{BB}/f_a-C_{WW}/f_a$ parameter space for fixed values of $C_{\mu\mu}$ and $m_a$.  
In Fig.~\ref{fig:general1}, we choose ALP masses $m_a = 5,~10,~50,~70$ and $100$ GeV. For muon coupling, we choose small value of $C_{\mu\mu}/f_a=-5$ and~ $-50 ~\text{TeV}^{-1}$, respectively (for 10GeV case, $C_{\mu\mu}/f_a=-50~\text{TeV}^{-1}$ is fully excluded by CMS $4\mu$ search and we show $C_{\mu\mu}/f_a=-10~\text{TeV}^{-1}$ instead, and for 100GeV, we choose $C_{\mu\mu}/f_a=-10~\text{TeV}^{-1}$) which can evade most of the constraints based on the searches for muonic final states. 
These values fall into the left branch in Fig.~\ref{fig:parameter-g-2}, which correspond to the small $C_{\mu\mu}$ solutions.
For larger $C_{\mu\mu}$, the constraints from $3\gamma/2\gamma$ searches are less stringent, but the $Z$-factory search channel $\mu^+\mu^-\gamma$ could probe more parameter spaces, similar to the situation of $C_{WW}=0$. 
The $(g-2)_\mu$ bands (red) in Fig.~\ref{fig:general1} have negative slope, corresponding to approximately constant $C_{\gamma \gamma}$. This implies that the contribution to $(g-2)_\mu$ is dominated by $C_{\gamma \gamma}$ coupling and the contribution from $Z/W$ bosons is less important.

For general ALP-gauge couplings, the constraint from $Z$ width precision measurement, $\Gamma_{Z}^{\rm tot}$, should be taken into account. If $C_{WW} = 0$, this constraint is weaker than the one from exotic photon final state searches. 
The $Z$ total decay width measured by LEP is $\Gamma_Z^{{\rm tot}}=(2.495\pm0.0023)\text{GeV}$ \cite{ALEPH:2005ab}. It requires the BSM branching ratio $\text{BR}(Z\to \text{BSM})<0.0018$ at $95\%$ C.L. It strictly limits the large $C_{WW}$ values for small $m_a$, as shown in Fig.~\ref{fig:general1}. The projected Tera-$Z$ $\Gamma_Z^{tot}$ measurement could further reduce the uncertainty to around 25 keV~\cite{AlcarazMaestre:2021ssl,CEPCPhysicsStudyGroup:2022uwl,Bernardi:2022hny}. 
For $m_a\lesssim 5$ GeV, the flavor violating meson exotic decay induced by non-zero $C_{WW}$ strongly constrains the parameter space of ALP~\cite{Bauer:2021mvw}. This constraint is absent for $m_a > 5\ \text{GeV}$.

For $m_a = 5$ GeV, the most stringent existing constraints are the LEP-I ($Z \to 2\gamma$), L3 ($Z \to (\gamma\gamma)\gamma$) searches and the $\Gamma_Z$ measurement. There is still a large portion of $(g-2)_\mu$ band which is not constrained but will be fully covered by Tera-$Z \to a \gamma  \to (\mu^+ \mu^-) \gamma$ except a small line presenting $C_{\gamma Z}^\text{eff}  \approx 0$. 
This is a general feature for exotic $Z$ decay channel $Z \to a \gamma $.

For $m_a = 10$ GeV, more existing searches are relevant because the invariant mass is large enough to go above the di-muon threshold at the CMS $pp\to 4\mu$ search and the di-photon threshold at the L3 $Z\to 3\gamma$ search. The constraints from L3 $Z\to 3\gamma$ leaves a cross-shaped region open.  The region with a positive slope corresponds to $C_{\gamma Z}^\text{eff}  \approx 0$ which minimizes the production cross-section. There is also an unconstrained region with a negative slope since $C_{\gamma \gamma}^\text{eff}  \approx 0$ leads to vanishing branching ratio ${\rm BR}(a \to \gamma\gamma)$. The limits from LEP-I $(Z\to 2\gamma)$ and the Z factory ($Z \to a\gamma \to 3\gamma$)  searches share the same feature as those from L3. The former is weaker and less constraining,  while the latter is very strong leaving open only the two (very fine-tuned) directions with $C_{\gamma\gamma,\gamma Z}^{\text{eff}} = 0$. The CMS $pp \to 4\mu$ search leaves a region open close to $C_{\gamma Z}^\text{eff} \approx 0$, since we have already chosen a pretty small $C_{\gamma\gamma}$ which means a large $C_{\gamma Z}$ is necessary to the production of $a \mu^+\mu^- $ final state.
The $Z$-factory $Z \to a \gamma  \to (\mu^+ \mu^-) \gamma$ search shares similar feature that only the fine-tuned direction $C_{\gamma Z}^\text{eff} = 0$ is allowed. The green shaded region in the plot shows the projection for the $(\gamma\gamma)\gamma @$ 240 GeV channel. Since $e^+e^-\to a(\gamma\gamma)\gamma$ at $\sqrt{s}=240\text{GeV}$ is not close to the $Z$-pole, the projected coupling limit is much weaker than the $(\gamma\gamma)\gamma$ channel at the $Z$-pole, except in the fine-tuned case where $C_{\gamma Z}^\text{eff}\approx 0$. Additionally, for small masses such as $m_a = 10$ GeV, the branching ratio of $a\to\gamma\gamma$ is suppressed by $m_a$ compared to $a\to\mu\mu$. Furthermore, the decayed $\gamma$ from a light ALP has a lower chance of passing the energy selection criteria. Therefore, the $(\gamma\gamma)\gamma @$ 240 GeV channel has difficulty covering the $(g-2)_\mu$ region.

For $m_a = 50$ GeV,  new constraints from the  ATLAS $Z \to 3\gamma$ and 
the OPAL $ee \to 3\gamma$ searches become relevant. The former still leaves two fine-tuned directions open with $C_{\gamma\gamma, \gamma Z}^\text{eff} = 0$. The latter fully excludes the $C_{\gamma Z}^\text{eff} = 0$ direction since OPAL data were taken during runs at higher energy than $m_Z$. The signal does not vanish in $C_{\gamma Z}^\text{eff} = 0$ direction, because it can be generated via $C_{\gamma \gamma }$ alone, as shown in the second diagram of Fig.~\ref{fig:agamma-finalstates}. Future $ee\to (\gamma\gamma)\gamma$ search at 240 GeV could also push forward on $C_{\gamma Z}^\text{eff}=0$ direction limit. 

For $m_a = 50$ and $70$ GeV, the $Z$-factory $Z \to a \gamma  \to (\mu^+ \mu^-) \gamma$ search leaves an open region with smaller $C_{\mu \mu}$ due to the smaller branching ratio of $a \to \mu^+ \mu^-$. Similar to the case of $m_a = 10$ GeV, $Z$-factory ($Z \to a\gamma \to 3\gamma$) search is very powerful, leaving only two fine-tuned directions with $C_{\gamma\gamma}^\text{eff} = 0$ and $C_{\gamma Z}^\text{eff} =0$. 

For $m_a = 100$ GeV, the limits from on-shell $Z$ decay no longer apply. The OPAL $ee\to (\gamma \gamma)\gamma$ provides significant limits to the parameter space which covers the  $C_{\gamma Z}^\text{eff} = 0$ direction. We show the LHC photon fusion constraints ($\gamma \gamma {\rm F}$) from PbPb ion collision and pp collision which is complementary to OPAL limits. Since LHC $\gamma \gamma {\rm F}$ constraints show similar limits comparing with OPAL, we only show their limits for $m_a = 100$ GeV for better readability.
Future $ee\to (\gamma \gamma)\gamma$ search @ 240 GeV is shown as green shaded region, which can cover the entire red $(g-2)_\mu$ band.

To summarize, in comparison with the minimal case,  including the possibility of $C_{WW} \neq 0$ does not qualitatively change the picture. This is due to the fact that the contribution to  $(g-2)_\mu$  is dominated by $C_{\gamma\gamma} $ coupling. The contributions from $Z/W$ bosons are less relevant. With $C_{WW} \neq 0$, there are two new fine-tuned directions $C_{\gamma\gamma,\gamma Z}^{\text{eff}} = 0$, which help to evade limits from the exotic $Z$ decays to photon final states. The $Z$-factory constraints are so powerful that the allowed points lie very close to a line with almost exact cancellation. Interestingly, the OPAL $ee\to (\gamma \gamma )\gamma$ search does not rely on on-shell $Z$ production, thus it can exclude $C_{\gamma Z}^\text{eff} = 0$ direction. 
In addition, this process can be applied to future Higgs factories at FCC-ee and CEPC running at $\sqrt{s}=$240 GeV, which further excludes the parameter spaces for larger mass $m_a > m_Z$.
Comparing $Z \to  (\gamma \gamma )\gamma$ at future $Z$ factories and $ee\to (\gamma \gamma )\gamma$ at Higgs factories, the number of signal events for the former are larger than the latter by about a factor of 100. Therefore, for $m_a < m_Z$, the limits are dominated by exotic $Z$ decay except for mass threshold and fine-tuned directions.

\section{Conclusions}
\label{sec:conclusion}
The deviation between $(g-2)_\mu$ measurement and SM prediction is a tantalizing sign of possible new physics beyond the Standard Model. A generic ALP with sizable couplings to muons and photons can provide a potential explanation for the $(g-2)_\mu$ anomaly. In this paper we start with ALP effective Lagrangian with general couplings to electroweak gauge fields and muons. We provide full $(g-2)_\mu$ calculations in ALP-fermion derivative coupling basis up to the 2-loop Barr-Zee diagrams. 
The importance of the inner loop counter term is addressed  and we also provide an understanding of the existence of the inner loop counter term with chiral transformation in Eq.~(\ref{eq:operator-change}). We also extended the previous results by including heavy gauge boson $Z$ and $W$ into the calculation up to the 2-loop level, which is necessary due to the existence of counter terms and the heavy axion mass.

To recast the existing search results to our scenario, we performed simulations. In comparison with most of the previous studies, our scenario has the coupling between the ALP to both the muon and the gauge bosons. Some unique channels open up for this scenario. One such example is  $Z \to \mu^+\mu^- \gamma$. In addition, we have updated the constraints from light particle searches, such as the CMS muonic force search \cite{CMS:2018yxg}, which can provide limits on the ALP model which has not been taken into account previously. Moreover, there are more Feynman diagrams involved in our scenario, the results in general can not be obtained by simple rescaling and signal simulation is necessary. We conclude that the existing results from experimental searches can place bounds on the parameter space relevant for an explanation of $(g-2)_\mu$ anomaly, but there is still a large open space.

 As a powerful experimental probe of the ALP scenario, we propose to search for $ (\gamma \gamma)\gamma$, $(\mu^+ \mu^-)\gamma$ and $(\mu^+ \mu^-)\mu^+ \mu^-$ at future Tera-$Z$ and Higgs factories, such as the CEPC and FCC-ee. They can be sensitive to parameter space relevant for an $(g-2)_\mu$ explanation up to  $m_a \sim 85$ GeV from exotic $Z$ decay searches, and the reach can be extended to $\sim 160$ GeV at future Higgs factories, which
are complimentary to other existing experimental searches. We also include the $C_{WW}$ coupling which is not considered before. In this case, the ratio $C_{\gamma Z}/C_{\gamma \gamma}$ is no longer fixed. Moreover, one of the couplings $C_{\gamma Z}, C_{\gamma\gamma}$ can become vanishingly small which can help to evade the existing bounds and even the future $Z$-factories probes. One exception is the  OPAL $(\gamma \gamma) \gamma$ search at energies between $181$--$209$ GeV \cite{OPAL:2002vhf} which can constrain the fine-tuned direction $C_{\gamma Z} =0 $. Future Higgs factories such as FCC-ee and CEPC running at 240 GeV can further exclude the parameter space up to $m_a \sim 160$ GeV through  $ (\gamma \gamma)\gamma$ and $(\mu^+ \mu^-)\gamma$ channels.

\begin{acknowledgements}
We would like to thank JiJi Fan for the helpful discussions. The work of JL is supported by the National Science Foundation of China under Grant No.12075005, 12235001, and by Peking University under startup Grant No.7101502458. The work of XPW is supported by the National Science Foundation of China under Grant No.12005009. LTW is supported by the DOE grant DE-SC0013642.
\end{acknowledgements}

\bibliographystyle{utphys}
\bibliography{ref}

\end{document}